\documentclass[twocolumn,twocolappendix,times]{aastex631}

\usepackage{bm}
\usepackage{breqn}

\newcommand\gsim{\lower.6ex\hbox{$\buildrel>\over\sim$}}
\newcommand\lsim{\lower.6ex\hbox{$\buildrel<\over\sim$}}

\shorttitle{ PKS~2131$-$021: A Unique SMBHB Candidate}
\shortauthors{O'Neill et al.}

\graphicspath{{./}{figures/}}

\begin{document}

\NewPageAfterKeywords 

\title{ The Unanticipated Phenomenology of the Blazar  PKS~2131$-$021: A Unique Super-Massive Black Hole  Binary  Candidate}

\correspondingauthor{Anthony Readhead}
\email{acr@caltech.edu}

\author{S. O'Neill}
\affiliation{Owens Valley Radio Observatory, California Institute of Technology,  Pasadena, CA 91125, USA}
\author{S. Kiehlmann}
\affiliation{Department of Physics and Institute of Theoretical and Computational Physics, University of Crete, 71003 Heraklion, Greece}
\affiliation{Institute of Astrophysics, Foundation for Research and Technology-Hellas, GR-71110 Heraklion, Greece}
\author{A.C.S. Readhead}
\affiliation{Owens Valley Radio Observatory, California Institute of Technology,  Pasadena, CA 91125, USA}
\affiliation{Institute of Astrophysics, Foundation for Research and Technology-Hellas, GR-71110 Heraklion, Greece}
\author{M.F. Aller}
\affiliation{2 Department of Astronomy, University of Michigan, 323 West Hall, 1085 S. University Avenue, Ann Arbor, MI 48109, USA}
\author{R. D. Blandford}
\affiliation{Kavli Institute for Particle Astrophysics and Cosmology, Department of Physics,20
Stanford University, Stanford, CA 94305, USA}
\author{I. Liodakis}
\affiliation{Finnish Center for Astronomy with ESO, University of Turku, Vesilinnantie 5, FI-20014, Finland}
\author{M.L. Lister}
\affiliation{Department of Physics and Astronomy, Purdue University, 525 Northwestern Avenue, West Lafayette, IN 47907, USA}
\author{P. Mr{\'o}z}
\affiliation{Astronomical Observatory, University of Warsaw, Al. Ujazdowskie 4, 00-478 Warszawa, Poland}
\author{C. P. O'Dea}
\affiliation{Department of Physics and Astronomy, University of Manitoba, Winnipeg, MB R3T 2N2,  Canada}
\author{T. J. Pearson}
\affiliation{Owens Valley Radio Observatory, California Institute of Technology,  Pasadena, CA 91125, USA}
\author{V. Ravi}
\affiliation{Owens Valley Radio Observatory, California Institute of Technology,  Pasadena, CA 91125, USA} 
\author{M. Vallisneri}
\affiliation{Jet Propulsion Laboratory, California Institute of Technology, 4800 Oak Grove Drive, Pasadena, CA 91109, USA}
\author{K.A. Cleary}
\affiliation{Owens Valley Radio Observatory, California Institute of Technology,  Pasadena, CA 91125, USA} 
\author{M. J. Graham}
\affiliation{Division of Physics, Mathematics, and Astronomy, California Institute of Technology, Pasadena, CA 91125, USA}
\author{K.J.B. Grainge}
\affiliation{Jodrell Bank Centre for Astrophysics, University of Manchester, Oxford Road, Manchester M13 9PL, UK}
\author{M.W. Hodges}
\affiliation{Owens Valley Radio Observatory, California Institute of Technology,  Pasadena, CA 91125, USA}
\author{T. Hovatta}
\affiliation{Finnish Centre for Astronomy with ESO (FINCA), University of Turku, FI-20014 University of Turku, Finland}
\affiliation{Aalto University Mets\"ahovi Radio Observatory,  Mets\"ahovintie 114, 02540 Kylm\"al\"a, Finland}
\author{A. L\"ahteenm\"aki}
\affiliation{Aalto University Mets\"ahovi Radio Observatory,  Mets\"ahovintie 114, 02540 Kylm\"al\"a, Finland} 
\affiliation{Aalto University Department of Electronics and Nanoengineering, PO Box 15500, 00076 Aalto, Finland}
\author{J.W. Lamb}
\affiliation{Owens Valley Radio Observatory, California Institute of Technology,  Pasadena, CA 91125, USA}
\author{T. J. W. Lazio}
\affiliation{Jet Propulsion Laboratory, California Institute of Technology, 4800 Oak Grove Drive, Pasadena, CA 91109, USA}
\author{W. Max-Moerbeck} 
\affiliation{Departamento de Astronomía, Universidad de Chile, Camino El Observatorio 1515, Las Condes, Santiago, Chile}
\author{V. Pavlidou} 
\affiliation{Department of Physics and Institute of Theoretical and Computational Physics, University of Crete, 71003 Heraklion, Greece}
\affiliation{Institute of Astrophysics, Foundation for Research and Technology-Hellas, GR-71110 Heraklion, Greece}
\author{T. A. Prince}
\affiliation{Division of Physics, Mathematics, and Astronomy, California Institute of Technology, Pasadena, CA 91125, USA}
\author{R.A. Reeves}
\affiliation{CePIA, Astronomy Department, Universidad de Concepci\'on,  Casilla 160-C, Concepci\'on, Chile} 
\author{M. Tornikoski}
\affiliation{Aalto University Mets\"ahovi Radio Observatory,  Mets\"ahovintie 114, 02540 Kylm\"al\"a, Finland}
\author{P. Vergara de la Parra}
\affiliation{CePIA, Astronomy Department, Universidad de Concepci\'on,  Casilla 160-C, Concepci\'on, Chile} 
\author{J. A. Zensus}
\affiliation{Max-Planck-Institut f\"ur Radioastronomie, Auf dem H\"ugel 69, D-53121 Bonn, Germany}

\begin{abstract}
Most large galaxies host supermassive black holes in their nuclei and are subject to mergers, which can produce a supermassive black hole binary (SMBHB), and hence periodic signatures due to orbital motion.  We report unique periodic radio flux density  variations in the blazar PKS~2131$-$021, which strongly suggest an SMBHB with an orbital separation of $\sim 0.001-0.01$ pc.   Our  45.1-year radio light curve shows two epochs of strong  sinusoidal variation with the same period and phase to within $\lsim2\%$ and $\sim 10\%$, respectively, straddling a 20-year period when this variation was absent. Our simulated light curves accurately reproduce the ``red noise'' of this object, and   Lomb-Scargle, weighted wavelet Z-transform, and least-squares sine wave analyses demonstrate conclusively, at the $4.6\sigma$ significance level, that the periodicity in this object is not due to random fluctuations in flux density.    The observed period  translates to $2.082\pm 0.003$  years in the rest frame at the $z=1.285$ redshift of PKS~2131$-$021.   The periodic variation in PKS~2131$-$021 is remarkably sinusoidal. We present a model in which  orbital motion, combined with the strong Doppler boosting  of the approaching relativistic jet, produces a sine-wave modulation in the flux density which easily fits the observations.  Given the rapidly-developing field of gravitational wave experiments with pulsar timing arrays, closer counterparts to  PKS~2131$-$021 and searches using the techniques we have developed are strongly motivated. These results constitute a compelling demonstration that the phenomenology, not the theory, must provide the lead in this field.

\end{abstract}


\section{Introduction}\label{sec:intro}

The identification of supermassive black hole binaries (SMBHBs) will open the field to multimessenger astronomy through the gravitational radiation they produce.  Pulsar timing arrays provide a powerful technique for searching for nanohertz signals from gravitational waves from SMBHBs through the timing of millisecond pulsars    \citep{2018MNRAS.481L..74H,2019A&ARv..27....5B}.    However, in spite of the fact that galaxy mergers are not uncommon, there are relatively few instances of two galaxies with supermassive black holes in their nuclei being seen in the actual process of the merging of the spheres of influence of their SMBHs, or of the following stages, when an SMBHB forms by ejecting stars from the merging central clusters, and spirals in more closely due to gravitational radiation, and finally coalesces    \citep{1980Natur.287..307B}.
A particularly fine example of the early stage of possible evolution towards an SMBHB is that of 3C~75    \citep{1985ApJ...294L..85O}, where both SMBHs are producing radio jets, and their projected separation  is 7.2 kpc. On pc scales the best  SMBHB candidate is B3\,0402+379    \citep{2006ApJ...646...49R, 2017ApJ...843...14B}, with a projected separation of 7.3 pc, a deduced period of $3 \times 10^4$yr, and a deduced SMBHB mass of $\approx 1.5\times 10^{10} M_\odot$. The strongest SMBHB candidate with a separation of $\ll 1$pc is OJ 287   \citep{1988ApJ...325..628S,2016ApJ...819L..37V,2021MNRAS.503.4400D}, for which the separation $\sim 0.1$ pc, the deduced primary mass $\approx 1.8\times 10^{10} M_\odot$, and the deduced secondary mass  $\approx 1.5   \times 10^8 M_\odot$. At  separations $\ll 1$pc, even with high-frequency very long baseline interferometry (VLBI), for all but the closest active galactic nuclei (AGN), we lack the angular resolution required to demonstrate the existence of an SMBHB through imaging, and we have to look for other signatures.

In principle, light curves offer a way forward    \citep{2009ApJ...700.1952H,2019A&ARv..27....5B}, because SMBHBs may reasonably be expected to exhibit periodicities. However, it has been pointed out    \citep{2016MNRAS.461.3145V,2019MNRAS.482.1270C} that, notwithstanding the rich literature on periodicities and quasi-periodic oscillations (QPOs) in blazars going back over five decades, there are very few statistically solid results. In their detailed analysis of ten blazars in which QPOs have been reported,   \cite{2019MNRAS.482.1270C} show that no strong cases for $\sim$year-long periodicities can be confirmed.  They are all consistent with the power spectra of the variations in these objects. It requires careful modelling  of the red noise in the power spectrum of a blazar to evaluate the significance of any claimed periodicity. \cite{2017A&A...600A.132S,2018A&A...615A.118S} had estimated that $\sim$10\% of bright $\gamma$-ray blazars are QPOs, but after their detailed analysis they withdrew this estimate \citep{2019MNRAS.482.1270C}.

In a search for strong periodic signals showing at least 1.5 cycles in the optical light curves of 243,500 quasars, \cite{2015MNRAS.453.1562G} found 111 candidates, of which the strongest is PG~1302$-$102  \citep{2015Natur.518...74G}. In this object approximately sinusoidal variations have been seen over a span of $\sim$20 years.  However \cite{2016MNRAS.461.3145V} have challenged the SMBHB interpretation of the PG~1302$-$102 optical light curve, attributing the periodicity  to red noise.  

 Since  blazar light curves at radio wavelengths have a red noise spectrum with a non-Gaussian probability density function (PDF) \citep{2017MNRAS.467.4565L}, except where stated otherwise, we assume a red noise PDF throughout this paper.

Given the history and the persisting problems, it is clear that 
great caution is  needed in the identification of periodicities and quasi-periodicities in active galactic nuclei. For this reason we regarded the   $P_\oplus=4.69{\rm yr}\pm0.14$ yr earth-frame observed periodicity reported in the blazar PKS~2131$-$021 by   \cite{2021MNRAS.506.3791R} as an interesting result to follow, but by no means yet shown to be a strong QPO or  SMBHB candidate.  The \cite{2021MNRAS.506.3791R} paper is based entirely on 11 years  (2008-2019) of our own 15 GHz monitoring observations of PKS~2131$-$021  with  the Owens Valley Radio Observatory (OVRO) 40~m Telescope.

We recently came across observations of PKS~2131$-$021   \citep{1986AJ.....92.1262O} made at the Haystack Observatory between 1975 and 1983, which show the same periodicity to $\lsim 2\%$, and  phase to within 10\% of the period.   As we show with extensive tests in this paper, the level of significance of this periodicity, when 45.1 years of radio monitoring data are combined, is $ 4.6\sigma$, and it is certainly not a red noise phenomenon. This makes PKS~2131$-$021 a strong QPO or SMBHB candidate.   In addition to the Haystack data, we have also added the 14.5 GHz light curve of the University of Michigan Radio Astronomy Observatory (UMRAO), which covers the period 1980-2012, and is in excellent agreement with the Haystack and OVRO light curves in the regions of overlap, thereby giving us an uninterrupted, well-sampled, 45.1-year 14.5 GHz - 15.5 GHz light curve of PKS~2131$-$021.

All of the data presented in this paper are from targeted observations of PKS~2131-021, i.e. they are not serendipitous observations in which PKS~2131$-$031 was observed in the fields of other objects. Given the inhomogeneous processing procedures, the diverse sets of flux calibrators, and the observed matched flux densities in the overlapping regions, it is clear that the unusual light curve of PKS~2131$-$021 is not a result of faulty processing. These demonstrate the basic light-curve integrity.

At the start of this project we confirmed four predictions related to the  periodicity seen in the OVRO data.  This convinced us that the periodicity is telling us something important about the physics of this object and is not simply a random variation due to red noise. We began with a least squares sine wave fit to the OVRO data, on the basis of which we predicted that the sine wave oscillations had begun before the OVRO observations, so we extrapolated the sine wave backwards and began a search for earlier data. Our prediction was that we would find earlier sinusoidal variations in phase with the OVRO observations. The first confirmed prediction came when we looked at MOJAVE \citep{2018ApJS..234...12L} and UMRAO  data  going back to 1995 and found an in-phase peak in 2005,  immediately preceding the first OVRO peak.  We subsequently obtained UMRAO data going back to 1980 and we found a second in-phase peak in 1982. This was the second confirmed prediction. The peak in 1982 was much larger than the peaks from 2005 to 2020, and we predicted that had we been observing prior to 1982 we would have seen a larger sinusoidal oscillation prior to 1982.  At the time we thought there were no data on PKS 2131-021 earlier than the UMRAO data.  However, through a literature search, we discovered the Haystack data, which began in 1975. This did indeed show an earlier in-phase peak, in 1976. This was the third confirmed prediction.  Furthermore, as predicted, like the peak in 1982, the peak in 1976 was  much larger than the sinusoidal variations from 2005 to 2020. This was the fourth confirmed prediction.  The chain of events occurred exactly as described here, and convinced us of the significance of the OVRO periodicity, which, as we show in this paper, has been confirmed by rigorous statistical analysis.

There are clearly several possible explanations for the periodicities observed in PKS~2131$-$021. Precession of the relativistic jet due to misalignment of the spin axis of the SMBH and accretion disk \citep{2004ApJ...616L..99C} is one possibility. Alternatively, the periodicity could be the result of precession due to misalignment of the orbital plane of a SMBHB with the accretion disk of the more massive SMBH \citep{2017ApJ...851L..39C}. Another possibility is precession due to warping of the accretion disk \citep{2018MNRAS.478.3199B, 2018NatAs...2..443A}. While these may all be viable explanations of the periodicity we see in PKS~2131$-$021, there is a more straightforward explanation -- namely that the periodicity is simply due to the orbital motion of the SMBHB.  We show here that all the observations can be explained by this simple model, and that no precession is needed to explain the light curve of PKS~2131$-$021, although, as we show, it might well explain the large-scale morphology.  It should be noted that, given the characteristic time-scales for variability in blazars, which range from months to years, and which are likely dominated by fuelling of the central engine, and given the multiple sites of radio emission along the jets, it is not difficult to invent models in which the sinusoidal signal switches on and off. Thus the appearance and disappearance of the sinusoidal variability is easily accommodated in any model, and for this reason we do not discuss it further in this paper. Since the SMBHB orbital motion explanation is the simplest, we apply Occam's Razor.  We deliberately do not consider other possible explanations in this paper, since it seems to us that nature is pointing the way.  Adopting the simplest explanation is the best way to proceed at this early stage in our understanding of the phenomenology of SMBHBs with relativistic jets.  Simple orbital motion was suggested as an explanation of blazar periodicities by \cite{2017MNRAS.465..161S}, but their model is a complex one, which does not produce sinusoidal variations in all circumstances, and not at all unless the jet itself is assumed to consist of a fast-moving ``spine'' surrounded by a slower-moving ``sheath''.  While this might well be the case in PKS~2131$-$021, it is  not required by our model, which is the simplest possible SMBHB-relativistic jet  model.

We wish, therefore, to make clear at the outset that in this paper we deliberately focus on a particular model, to the exclusion of other viable models, because the phenomenology of the sinusoidal flux density variations -- which has not been anticipated in previous studies -- is an inevitable~and inescapable consequence of the SMBHB orbital motion of the relativistic jet, and we are strongly of the opinion that we should pursue this approach until the phenomenology {\it requires} more complex models.

In this paper we consider all periodicities with significance levels below  $3\sigma$ to be red noise, unless other, uncorrelated and independent, observations raise the significance to the $\ge 3\sigma$ level.
We analyze the PKS~2131$-$021 light curve and show that it is unique amongst AGN that have been considered as possible SMBHB candidates, and is indeed  a prime SMBHB candidate. In \S~\ref{sec:observations} we describe the observations; in \S~\ref{sec:analysis} we analyze the light curve using three different approaches and taking great care to model the red noise correctly and hence to derive robust measures of the significance of our results,  and derive an upper limit to the chirp mass of the putative SMBHB based on the radio observations alone;  in \S~\ref{sec:model} we present  a model of PKS~2131$-$021, in which the observed periodicity is the orbital period of the putative black hole binary, which can account for the sinusoidal shape and amplitude of the periodic variability that we see; in \S~\ref{sec:nanograv} we discuss the expected gravitational wave strain and derive an upper limit on the chirp mass of the SMBHB based on the upper limits derived from pulsar NANOGrav observations  \citep{2019ApJ...880..116A}.

The redshift of PKS~2131$-$021 is $z = 1.285$  \citep{1997MNRAS.284...85D,2001AJ....122..565R,2006AJ....132....1S}, which we have recently confirmed (see \S~\ref{sec:optical}).
For consistency with our other papers, we assume the following cosmological parameters: $H_0 = 71$\, km\,s$^{-1}$\, Mpc$^{-1}$, $\Omega_{\rm m} = 0.27$, $\Omega_\Lambda = 0.73$  \citep{2009ApJS..180..330K}.  On this model the comoving coordinate distance of PKS~2131$-$021 is 3.97~Gpc, the angular diameter distance is 1.74~Gpc, and the luminosity distance is 9.08~Gpc. None of the conclusions would be changed were we to adopt the best model of the Planck Collaboration  \citep{2020A&A...641A...6P}. 

\begin{figure*}[ht!]
\centering
\includegraphics[width=0.8\linewidth]{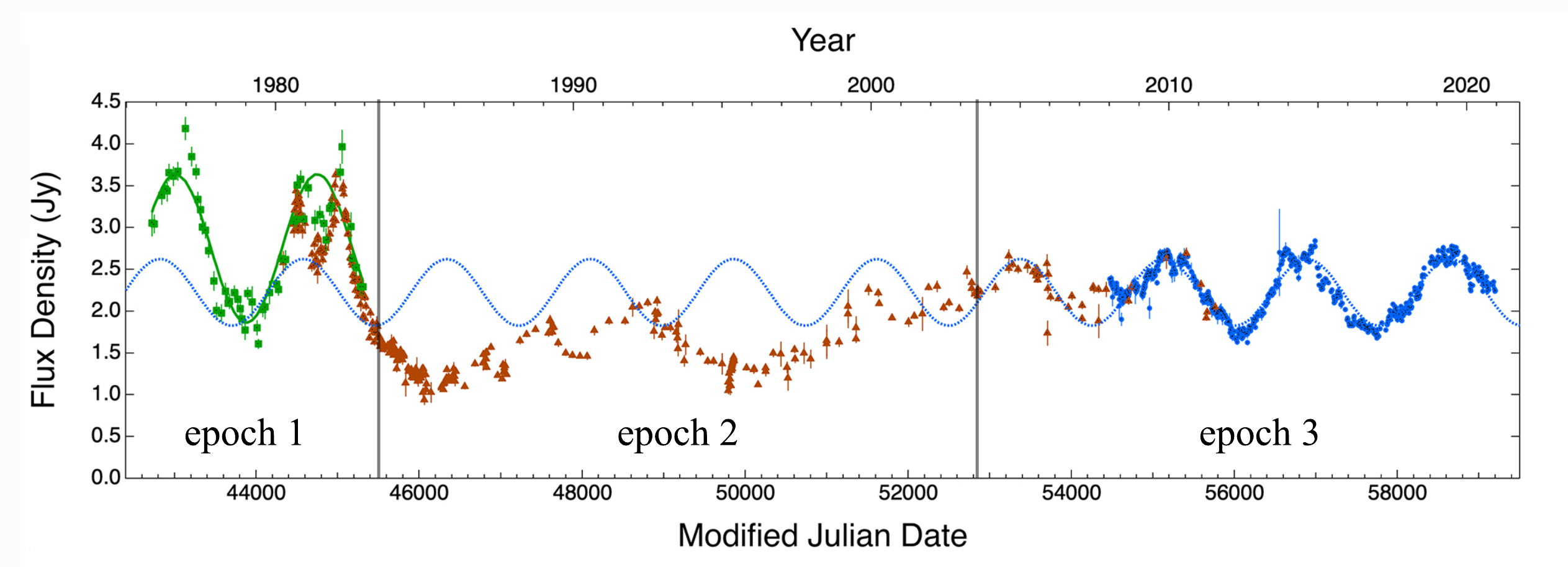}
\caption{The 14.5~GHz - 15.5~GHz light curve of PKS~2131$-$021. The Haystack data are shown by the green squares, the UMRAO data by the brown triangles and the OVRO data by the blue dots. Note the excellent agreement between Haystack and UMRAO and between UMRAO and OVRO in the regions of overlap.   The light curve shows three distinct epochs of activity: In epoch~1 (MJD$<$45500) there is a strong periodic signal, $P_1=1729.1\pm32.4$~days; in epoch~2 (45500$<$MJD$<$52850) this periodic signal is absent; in epoch~3 (MJD$>$52850) the periodic signal of epoch~1 returns, with $P_3=1760.4\pm5.3$~days, in phase but with lower amplitude than that of epoch~1.  The solid green line shows the least-squares sine wave fit to the Haystack data, and the dotted blue line shows the least-squares sine wave fit to the OVRO data  extrapolated backwards to provide a comparison with the Haystack data. The periods of the Haystack and OVRO sine wave fits match  to $\lsim 2\%$. The phase of the OVRO periodic signal extrapolated back to epoch~1 matches to $\sim 10\%$ of the period.}
\label{plt:light curves}
\end{figure*}

\section{The Observations}\label{sec:observations}
In this section we describe our radio monitoring and VLBI observations, and, in addition, infrared and optical observations of PKS~2131$-$021.  Note that for all the single dish observations the angular extent of PKS~2131$-$021 is $\ll$ the telescope beamwidth, so that the total flux density is always measured.

\subsection{The Radio Light Curve Observations}\label{sec:radio1}

The 45.1-year radio light curve of PKS~2131$-$021 is shown in Fig.~\ref{plt:light curves}.  It is immediately apparent to the eye that there are three distinct epochs, which are demarcated by the vertical black lines in Fig.~\ref{plt:light curves}: In epoch~1 a strong 1.5-cycle  periodic oscillation is seen in the Haystack data (green squares), of which the last half-cycle  is also seen in the UMRAO data (brown triangles). There follows a 20-year period (epoch~2) in which the oscillation of epoch~1 is not seen.   In 2003 we enter the third epoch~in which a strong oscillation is again seen, that has very nearly the same period (within 2\%) and also the same phase  (within 10\% of a cycle) as the periodic oscillation of epoch~1.  This analysis by inspection is well corroborated by both the Lomb-Scargle (LS) \citep{1976Ap&SS..39..447L,1982ApJ...263..835S} and weighted wavelet Z-transform (WWZ) \citep{1996AJ....112.1709F} analyses we carried out (see \S~\ref{sec:analysis}).

\begin{figure*}[!t]
\centering
\includegraphics[width=0.8\linewidth]{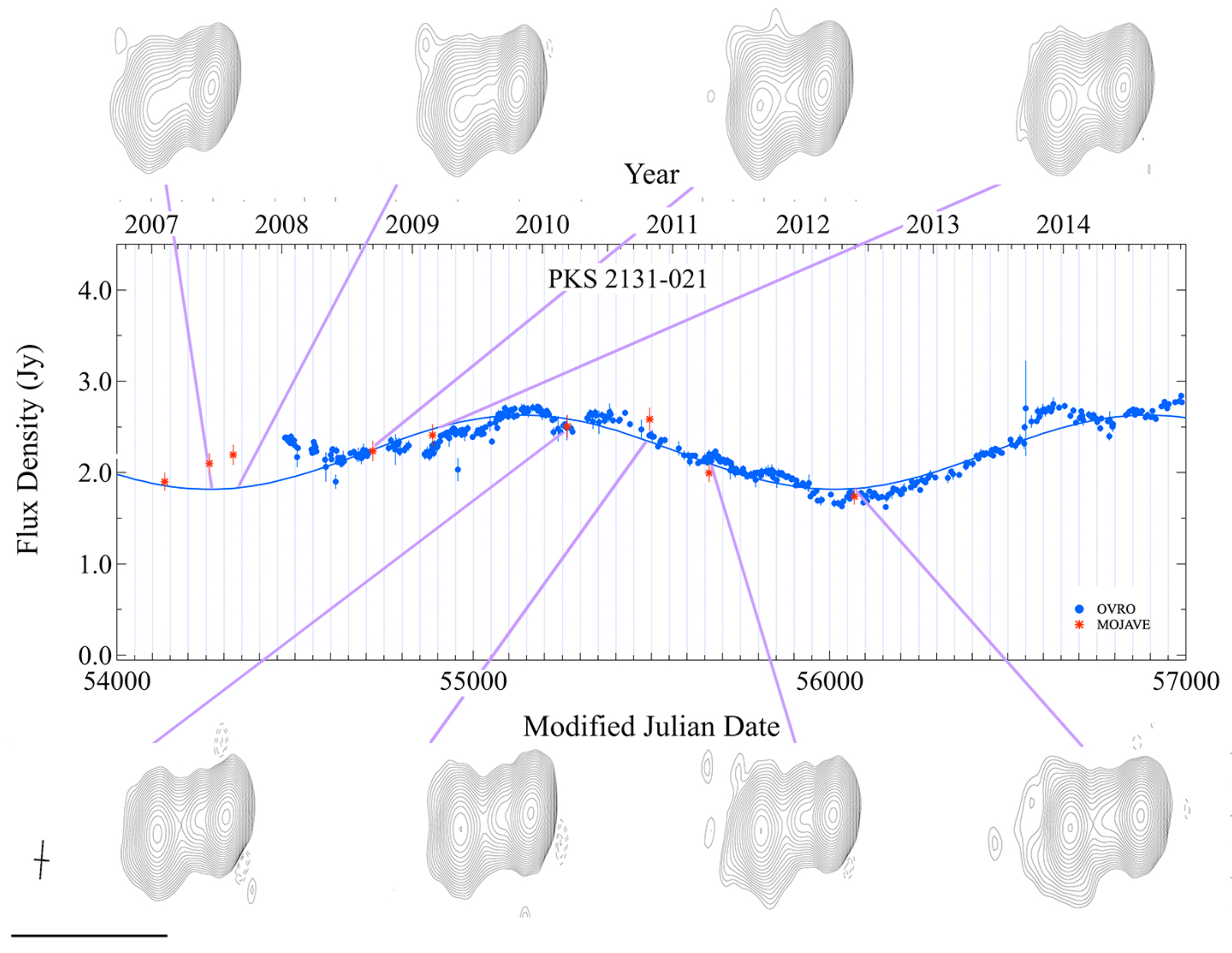}
\caption{MOJAVE 15 GHz VLBI maps \citep{2018ApJS..234...12L} matched to the OVRO 15 GHz light curve (blue dots) and least-squares sine wave fit to the OVRO data (blue line). The total flux densities measured by MOJAVE are shown by the red asterisks. The maps show the basic core-jet structure of PKS~2131$-$021 on pc scales. The bar at the lower left indicates the scale of 5 milliarcseconds  ($\sim 42$ pc)  of all the maps, and the small cross indicates the typical full width half maximum beam size and orientation. Detailed comparison of the maps, models, and light curve shows that it is the core (primarily) and innermost component (partly), at $\sim 0.3$ milliarcseconds from the core,   that are responsible for the periodic flux density variations.}
\label{plt:maps}
\end{figure*}

{\it Haystack Observations:} The 15.5~GHz ($\lambda$ 1.9 cm) data
\citep{1986AJ.....92.1262O} are shown by the green square data points in Fig.~\ref{plt:light curves}. The bandwidth was 1200 MHz and the primary beamwidth (FWHM) was 2.2 arc minutes. These data were obtained at  roughly one month intervals with the 120-ft diameter telescope of the Haystack Radio Observatory located in Westford, MA. Details of the observing procedures  \citep{1974AJ.....79.1232D,1982PhDT.........6B} are briefly summarized here. The observations were taken with the dual-feed beam-switched system which reduces the effect of atmospheric fluctuations. Pointing corrections were applied and the source antenna temperature was measured by an ``on-off'' procedure. Visual inspections of a chart record were used to identify  observations affected by interference, which were then deleted. Corrections for atmospheric extinction and elevation-dependent gain were applied. The primary flux density calibrators were the compact HII region DR21, 3C274, and 3C123.  A  correction of a few percent was applied to account for partial resolution of DR21 and 3C274 by the Haystack observing beam. 

\vskip 2pt
\noindent
{\it UMRAO Observations:} The UMRAO light curve is shown by the brown triangular data points in Fig.~\ref{plt:light curves}.  The long term 14.5~GHz  total flux density data for PKS~2131$-$021 were obtained with the 26-meter equatorially-mounted, prime-focus University of Michigan paraboloid as part of the UMRAO monitoring program which operated from the mid 1960s until 2012.5 and included observations of total flux density and linear polarization at three cm-band wavelengths for hundreds of blazars in a dedicated AGN program.  A description of the general observation and calibration procedures is given in  \cite{1985ApJS...59..513A}.  In brief, at 14.5~GHz dual, rotating linearly-polarized feed horns placed symmetrically about the prime focus were employed and alternated beams on the source; these fed a broadband uncooled high electron mobility
transistor (HEMT) amplifier with a bandwidth of 1.68~GHz.  Each observation shows the average from one day of a series of on-on measurements obtained by alternating between rotating feed horns during a 30-40~minute interval. The cadence was one of these averages per week.  The adopted flux density scale   \citep{1977A&A....61...99B} uses Cas A as the primary calibrator, accounting for its measured decay rate \citep{1974ApJ...188L..11D}. Observations of secondary calibrators, were interleaved approximately every 2 hours throughout each 14.5~GHz observing run to monitor pointing and gain changes. The standard deviation associated with each daily flux density observation is computed from the system noise temperature and the number of individual on-on measurements made on the particular day. The standard error estimates shown include the effects of measurement noise, the errors introduced by uncertainties in the pointing corrections applied to the observations, and the uncertainties in determining  the antenna gain as a function of time.
The goal of the UMRAO program was to detect and follow blazar flares, so in general the cadence of the observations was set by the variability state (high cadence during flaring). However, as a member of the UMRAO BL Lac sample \citep{1999ApJ...512..601A}, PKS~2131$-$021 was also routinely observed approximately every three months. Additionally, during the last decade of operation, quasi-simultaneous observations (within a week) were matched to MOJAVE epochs for calibration of the VLBA data.  

\vskip 2pt
\noindent
{\it OVRO Observations:} The OVRO light curve is shown by the blue dot data points in Fig.~\ref{plt:light curves}. The OVRO data were part of the OVRO 40~m Telescope Monitoring Program \citep{2011ApJS..194...29R}. The telescope uses off-axis dual-beam optics in which the beamwidth (FWHM) is 157 arc seconds. The cryogenic receiver uses a HEMT amplifier and is centered at 15~GHz with 2~GHz equivalent noise bandwidth. Gain fluctuations and atmospheric and ground contributions are removed with the double switching technique \citep{1989ApJ...346..566R} where one of the beams is always pointed on the source. Until May~2014 the two beams were rapidly alternated using a Dicke switch. In May~2014 a new pseudo-correlation receiver replaced the old receiver. Since then a 180~degree phase switch has been used to alternate the beams.
A temperature-stable~noise diode is used for relative calibration to compensate for gain drifts. The primary flux density calibrator is 3C~286 with an assumed value of 3.44~Jy \citep{1977A&A....61...99B}, and DR21 is used as a secondary calibrator source. Details of the observation and data reduction \citep{2011ApJS..194...29R} cover the absolute calibration and the uncertainties, which include both the thermal fluctuations in the receiver and systematic errors that have been added in accordance with a rigorous procedure \citep{2011ApJS..194...29R}.

\vskip 2pt
\noindent
{\it MRO Observations:} 
The 22 GHz and 37 GHz observations were made with the 13.7 m diameter Aalto University Mets\"ahovi radio telescope, which is a radome enclosed antenna  with Cassegrain optics  in Finland ($60\degr13' 04''$\,N, $24\degr23'35''$\, E). The receivers have HEMT front ends operating at room temperature. The bandwidth at both frequencies is 1 GHz, and the beamwidths (FWHM) are 4.0 arcminutes and 2.4 arcminutes at 22 GHz and 37 GHz, respectively. The observations are Dicke switched ON--ON observations, alternating the
source and the sky in each feed horn. A typical integration time to obtain one flux density data point is between 1200 and 1800 s. The detection limit of the system at 22/37 GHz is on the order of 0.2 Jy under optimal conditions, but is heavily weather-dependent.
The flux density scale is set by observations of the HII region DR 21. Sources NGC 7027, 3C~274 and 3C~84 are used as secondary calibrators. A detailed description of the data reduction and analysis is given in  \cite{1998A&AS..132..305T}. The error estimate in the flux density includes the contribution from the measurement RMS and the uncertainty of the absolute calibration.

\begin{figure}[ht!]
\centering
\includegraphics[width=1.0\linewidth]{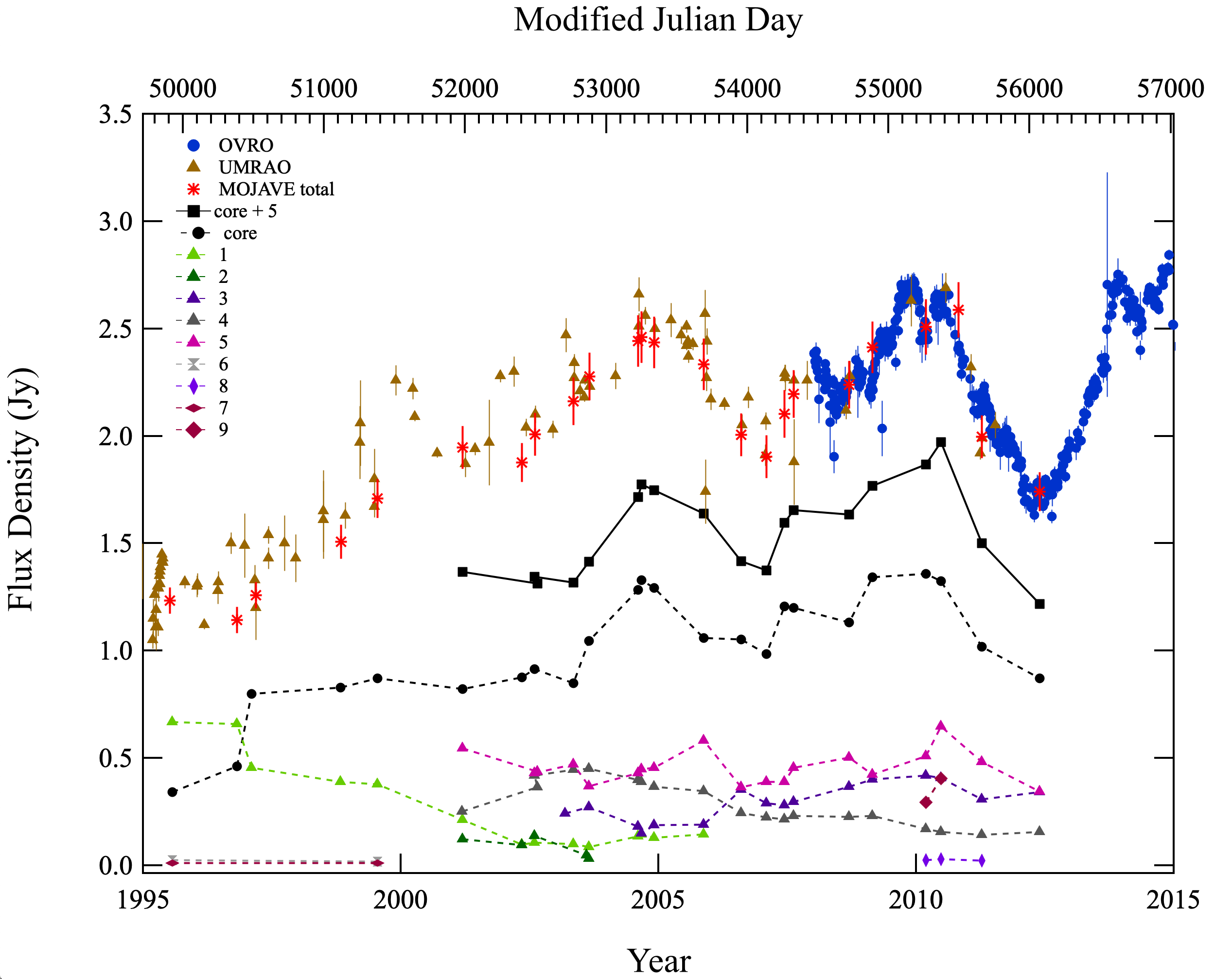}
\caption{Radio light curves of the entire source from UMRAO at 14.5~GHz (brown triangles), OVRO at 15.0~GHz (filled blue circles), and MOJAVE at 15.0 GHz (red asterisks), as well as individual pc-scale jet features as measured by the MOJAVE program. }
\label{plt:components}
\end{figure}

\begin{figure}[ht!]
\centering
\includegraphics[width=1.0\linewidth]{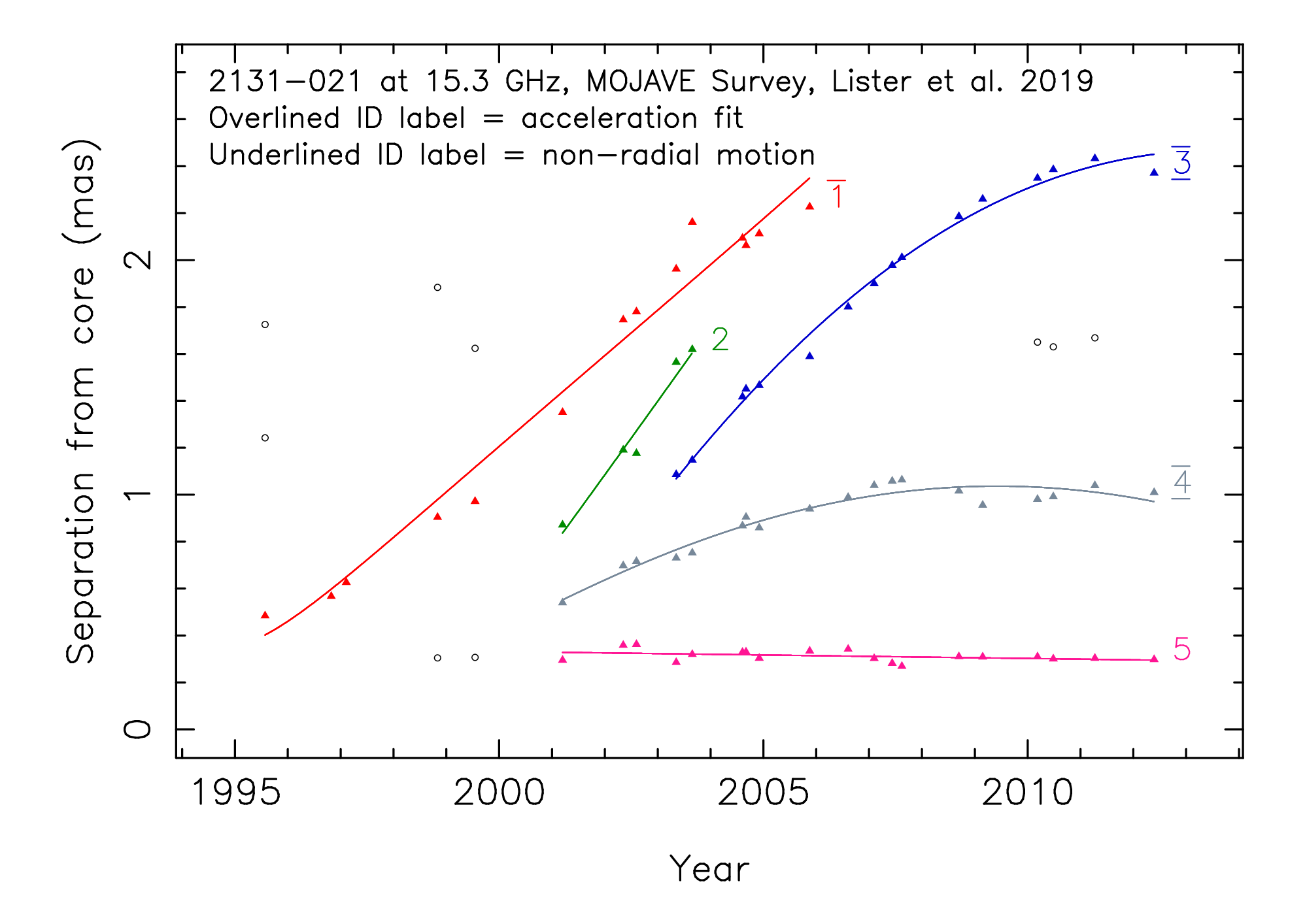}
\caption{The angular separation of jet components from
the core versus time for individual Gaussian model fitted features in
PKS~2131$-$021 on the MOJAVE program. Colored symbols
indicate robust features for which kinematic fits were obtained. The
identification number is overlined if an acceleration model was fitted
and indicated a $\ge 3\sigma$ acceleration. An underlined
identification number indicates a feature with non-radial motion. The black circles represent Gaussian components fitted to the jet emission that could not be clearly cross-identified over a minimum of five epochs.
}
\label{plt:separations}
\end{figure}

\subsection{15 GHz VLBI Observations}\label{sec:vlbi}

PKS~2131$-$021 was observed with the VLBA at 15~GHz on numerous
occasions between 1995 and 2012 as part of the 2~cm Survey  \citep{1998AJ....115.1295K} and MOJAVE  \citep{2018ApJS..234...12L}
programs. The last eight maps made on this program, six of which overlap the OVRO monitoring observations, are shown in Fig.~\ref{plt:maps}. These show the basic core-jet structure of PKS~2131$-$021 on pc scales. The results have been analyzed for 24 VLBA epochs \citep{2019ApJ...874...43L}  (including 6 from other projects obtained from the NRAO data archive) by fitting Gaussian
model features to the interferometric visibilities and tracking their
evolution in flux density (Fig.~\ref{plt:components}) and separation (Fig.~\ref{plt:separations}).  The VLBA images have a
restoring beam with typical FWHM dimensions $\sim$ 1 mas x 0.5 mas,
but the positional determination accuracy of the individual model features is
approximately 10 times better. The pc-scale radio morphology of the
source consists of a strong core feature with brightness temperature varying between 
$10^{10.6}$\,K and $10^{11.9}$\,K and a jet structure that extends
2.5 mas (21 pc projected) to the east. The core accounts for roughly half of the total 15~GHz flux  density of the source (Fig.~\ref{plt:components}). The jet and core are linearly polarized, with
fractional polarization levels increasing from 5\% to 15\% downstream.
These values are typical of other core-dominated blazars in the
MOJAVE sample \citep{2005AJ....130.1389L}.  Observations in 2006 \citep{2012AJ....144..105H}, showed
low Faraday rotation measure, with values ranging up to $131$ rad m$^{-2}$,
and the electric vectors are generally
aligned with the jet direction, indicative of a magnetic field
oriented perpendicular to the flow. The one exception was a jet
feature (component 3) that had electric vectors oriented in the transverse jet direction. 

In Fig.~\ref{plt:components} we show the light curves of the individual milliarcsecond jet components measured by MOJAVE  \citep{2019ApJ...874...43L}. These overlap the 2005 peak in the sine wave signal seen by UMRAO and the 2010 peak seen by both UMRAO and OVRO. Note that the core component and the component nearest to the core (component 5) are the components that produce the periodic signal.   The sum of these two components is shown by the solid line in Fig.~\ref{plt:components}.  In Fig.~\ref{plt:separations} we show the separations of the individual components from the core.  Component 5 is stationary ($\mu = 3.0\pm 1.9 \mu{\rm as}/$yr,  \citealp{2019ApJ...874...43L}) at a separation of 0.3 mas (0.2 pc) from the core. We note that, while we might expect all of the periodic variability to arise in the core, in view of the small separation of component 5 from the core, it is not unlikely that the core and component 5 will vary in phase on the timescale of the observed periodicity.

Based on near-simultaneous OVRO and VLBI observations, the VLBI flux density is $\gsim 98\%$  of the single dish flux density, so that, at most, $\sim $ 25 mJy of flux density is missing on small angular scales and is distributed on large scales.  PKS~2131$-$021 has the typical blazar jetted-AGN morphology of a one-sided jet with a flat spectrum core at one end of a steep spectrum jet. The components in the jet have been modelled  \citep{2019ApJ...874...43L} with the results shown in Figs. \ref{plt:components} and \ref{plt:separations}. The parsec-scale jet of PKS~2131$-$021 showed significant
activity during the period between mid 1993 and 2000, with several
features emerging from the core and moving downstream at apparent superluminal speeds.
No new moving features appeared in the jet between 2000 and 2013. 
As can be seen in  Fig.~\ref{plt:components}, at all epochs except the first, the core dominates the flux density.
From 15 March 2001 a stationary component (\#5) is seen at a distance of 0.3 milliarcseconds from the core,
and together the core plus component  5 dominate the light curve from then on. The sinusoidal flux density fluctuations seen in epoch~3 are clearly due to these two components, and dominated by the core.  Thus the periodicity we observe in PKS~2131$-$021 is coming primarily from unresolved components closer to the central engine and the base of the jet than the structures observed with 15 GHz VLBI.

\begin{figure*}[ht!]
\centering
\includegraphics[width=0.7\linewidth]{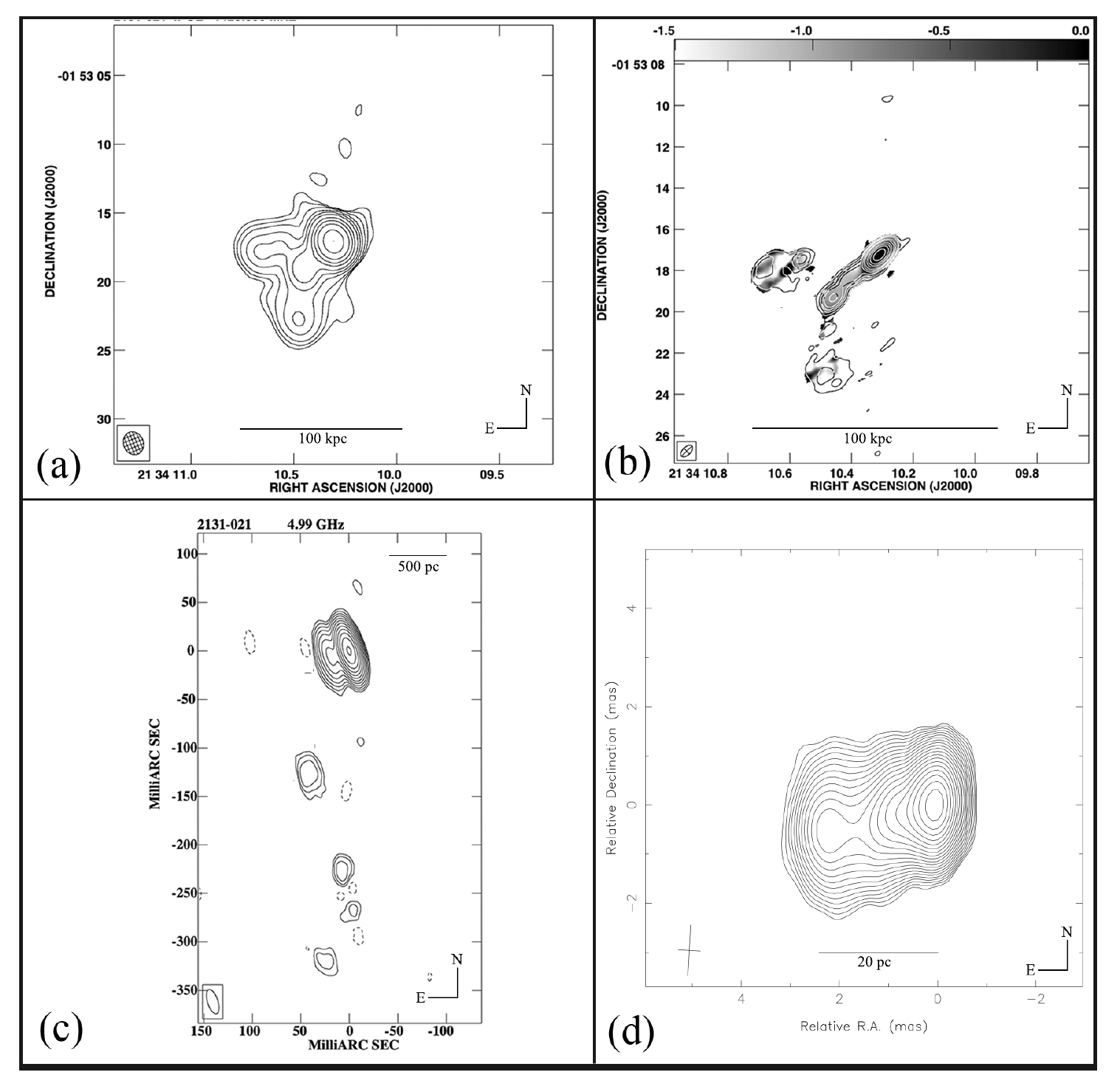}
\caption{Maps of PKS~2131$-$021 on different angular scales.  (a) VLA 1.4 GHz map from \cite{2001AJ....122..565R}. (b) VLA 4.86 GHz map from \cite{2003AJ....125.2447R}, the gray scale shows the 4.86 GHz - 8.46 GHz spectral index, which shows that the core has a flat spectrum and the jet has a steep spectrum. (c) EVN+MERLIN  4.99 GHz map from \cite{2002A&A...381..378C}. The RMS noise level is 0.2 mJy/beam and the peak flux density is 1349 mJy/beam. (d) Stack of 24 MOJAVE images \citep{2019ApJ...874...43L} from 1995-07-28  to 2012-05-24. Full width half maximum beamwidth: $1.04 \times 0.45$ milliarcseconds in position angle $-3^\circ$. Contours: $\sqrt{2} \times 1.714 $ mJy/beam.}
\label{plt:composite}
\end{figure*}

Kinematic analysis \citep{2019ApJ...874...43L} identified five
distinct features in the jet, each having a different apparent speed.
The innermost feature (id = 5) had no detectable~motion over a 12 year
period, while the speeds of the other features ranged from $2.6 c \pm
0.2 c$ (id = 4) to $20 c \pm 2.6 c$ (id = 2). Most of the moving
features display accelerated and/or non-radial motions on the sky that
are exaggerated by projection effects. Given the maximum apparent
speed, the jet viewing angle, $\theta$, is constrained to lie within
$5.7^\circ$ of the line of sight, and an analysis based on the median
core brightness temperature yields $\theta = 3.8^\circ$ and a Doppler factor
of $\delta = 14$ \citep{2021arXiv210904977H}.  

The exact dates of emergence of the individual moving features are difficult
to determine since that would require extrapolations of
unknown accelerations close to the core. Also, in the case of components
3 and 4, there is blending with other jet emission that made it
impossible to measure their positions when they were located near the core.
Simple constant velocity radial fits to components 1 and 2 give
ejection dates of $1993.7 \pm 0.4$ and $1998.6 \pm 0.5$ respectively.
For component 3, the first five epochs (when the feature was moving
approximately radially away from the core) imply an ejection
date of $1999.2 \pm 0.3$. In the case of the strongly accelerating component 4, 
its time of ejection can only be narrowed down to the interval 1997--2000.

\begin{figure*}[!t]
\centering
\includegraphics[width=0.9\textwidth]{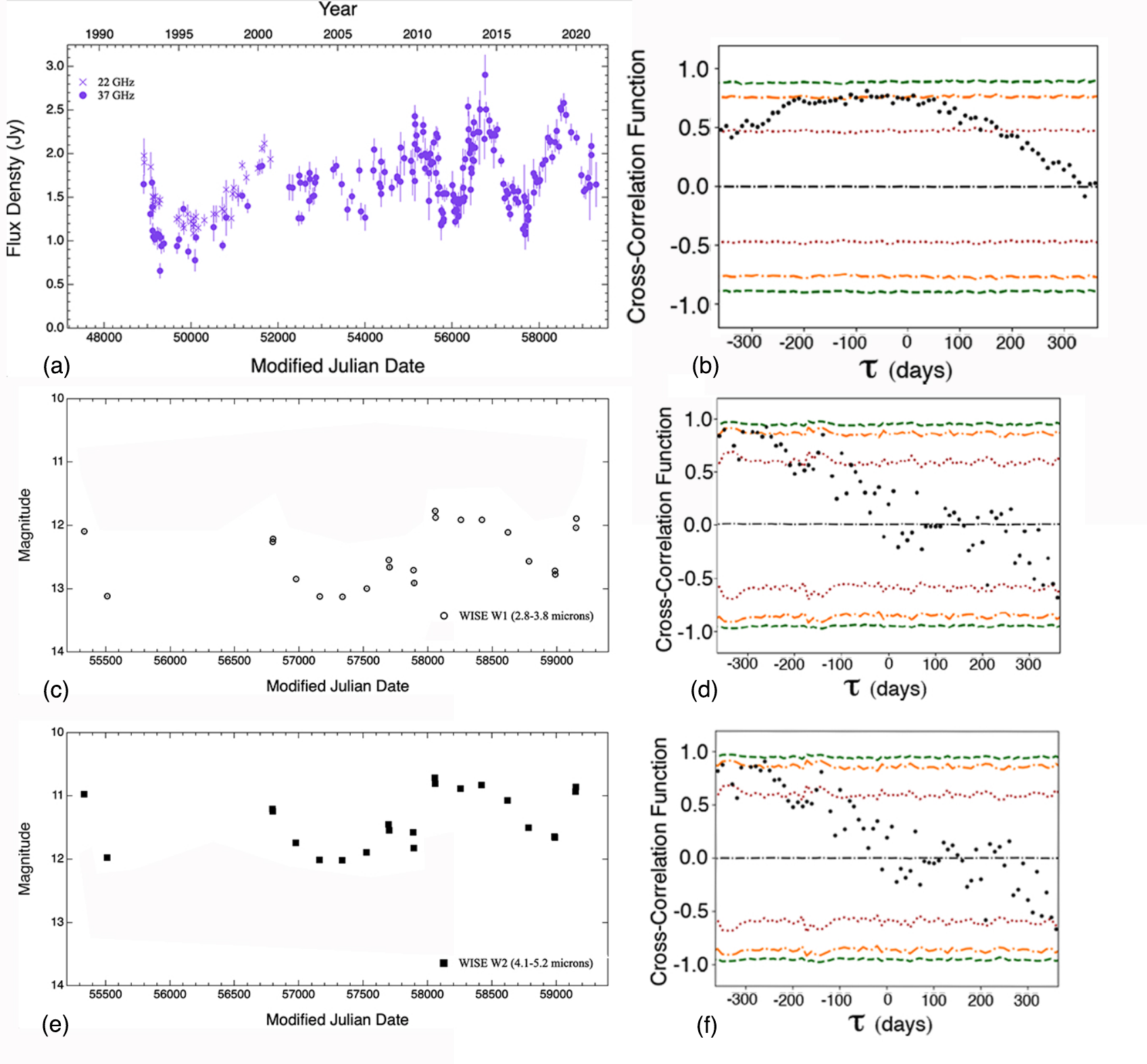}
\caption{PKS~2131$-$021 multifrequency light curves and cross-correlations. (a) light curves at 22 GHz (purple crosses), and 37 GHz (purple dots) from MRO. (b) Cross-correlation of the MRO 37 GHz and OVRO 15 GHz light curves. (c)  WISE infrared light curve in  the W1 ($2.8-3.8 \mu$m) band. (d) Cross-correlation of the 
WISE W1 (using flux density) and OVRO 15 GHz light curves. (e)  WISE infrared light curve in  the  W2 ($4.1-5.2 \mu$m band). (f) Cross-correlation of the WISE  W2 (using flux density) and OVRO 15 GHz light curves.  In both (c) and (e) the error bars are smaller than the symbols. The pink, orange, and green dashed curves indicate the $1\sigma$, $2\sigma$, and $3\sigma$ significance levels of the cross-correlation, using our simulated light curves, which assume a simple power law form  with a slope of 2, typical for blazars in these bands.}
\label{plt:multifreq}
\end{figure*}

\subsection{The Large Scale Radio Structure of PKS~2131-021}\label{sec:large}

The radio structure of PKS~2131$-$021 on different angular scales is shown in Fig.~\ref{plt:composite}. On arc-second scales PKS~2131$-$021 has an unusual structure, which is most clearly seen
in the 4.86 GHz VLA map of \cite{2003AJ....125.2447R} - reproduced  here in Fig.~\ref{plt:composite}(b),   which shows two linear features
extending to about 8 arc seconds from one side of the nucleus, and nothing on the opposite side of the nucleus.
An EVN-MERLIN map by \cite{2002A&A...381..378C} shows this bifurcated structure extending inwards to  within 100 milliarcseconds of the core.  Both gravitational lensing and precession have been suggested as possible causes for the strange structure. Another possibility, which has not previously been suggested to the best of our knowledge, is twin jets, such as are seen in 3C~75 \citep{1985ApJ...294L..85O}. It is interesting that two of the possible explanations for the large-scale structure of PKS~2131$-$021 (precession and twin jets) could well require  an SMBHB.

\subsection{Higher Frequency Radio Observations}\label{sec:higherfreq}

The MRO 22 GHz and 37 GHz observations are shown in Fig.~\ref{plt:multifreq}(a).
We have cross-correlated the MRO 37 GHz observations with the OVRO 15 GHz light curve, with the results shown in  Fig.~\ref{plt:multifreq}(b). There is clearly a correlation, as can also be seen by visual examination of the light curves, and it also appears that the 15 GHz variations lag those at 37 GHz, but it is hard to determine the lag -- it could be anywhere between $\sim 0$ and $\sim$ 200~days.

\subsection{Infrared Observations}\label{sec:infrared}
We have extracted the Wide Field Infrared Explorer (WISE) data in the 2.8--3.8 micron and 4.1--5.2 micron bands. These are shown in Fig.~\ref{plt:multifreq}(c) and (e). The cross-correlation between the flux densities of these two bands with the radio light curve can be seen in  Figs.~\ref{plt:multifreq}(d) and (f). The cross-correlation function is rather noisy but it likely shows a real correlation when it is at, or just above, the $2\sigma$ level. If this is to be believed then the radio variations lag the infrared variation by $\sim 250$--$350$~days.

\subsection{Optical Observations}\label{sec:optical}

We observed PKS~2131$-$021 on 2021 October 06 (HST) with the Low Resolution Imaging Spectrometer (LRIS) mounted on the Keck I telescope. The observations were carried out in dark time, under photometric conditions with a median seeing disk FWHM of 0.5\,arcsec. A 900\,s exposure was obtained at an airmass of 1.08, with the D560 dichroic, the 400/3400 blue-side grism, and the 400/8500 red-side grating (central wavelength 7830\,Angstrom). Spectral flux and telluric-absorption calibration was carried out using an observation of the standard star Feige 34. Data reduction followed standard procedures using the lpipe software  \citep{perley2019}. The software was used to perform bias subtraction using the overscan levels, flat-fielding using dome-flat exposures, automatic cosmic-ray rejection, wavelength calibration using internal comparison-lamp exposures and sky emission lines, optimal sky-line subtraction, and optimal extraction of the spectral trace. 
No significant difference was observed between our spectrum of PKS~2131$-$021 and a previous spectrum obtained on 1995 Nov 20  \citep{2001AJ....122..565R}. We confirm the redshift of 1.285, together with the BL Lac nature of the object.

\begin{figure}[!h]
\centering
\includegraphics[width=1.0\linewidth]{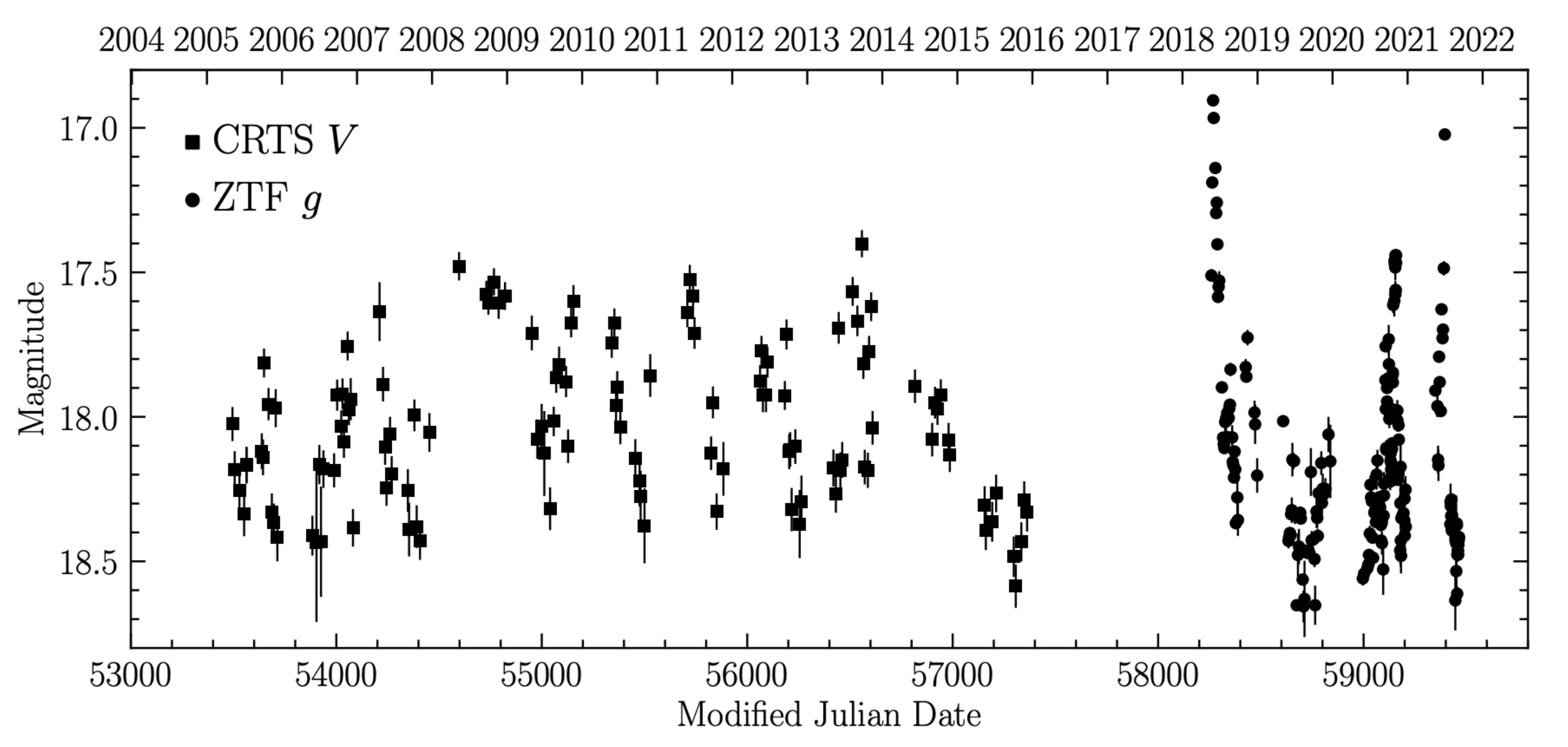}
\caption{Archival optical light curve of PKS\,2131$-$021.}
\label{fig:optical_lc}
\end{figure}

The archival optical light curve of the blazar is presented in Figure~\ref{fig:optical_lc}. It consists of 119 $V$-band epochs from the Catalina Real-Time Transient Survey (CRTS)  \citep{drake2009} and 203 $g$-band epochs from the Zwicky Transient Facility (ZTF)  \citep{masci2019,graham2019}. The CRTS data were collected between May 2005 and December 2015 using the 0.7\,m Catalina Schmidt Telescope located north of Tucson, Arizona. The ZTF data were taken between May 2018 and September 2021 with a wide-field camera mounted on the Palomar 48-inch Oschin (Schmidt) telescope.

We searched for periodic variations in the optical data. We ran generalized Lomb Scargle (GLS) \citep{1976Ap&SS..39..447L,1982ApJ...263..835S} and analysis of variance  period-searching algorithms \citep{czerny1989} on the CRTS and ZTF data and we did not find any significant periodicity in the range 1--1850~days. The optical light curve is not correlated with the radio data.

\begin{figure}[!b]
\centering
\includegraphics[width=1.0\linewidth]{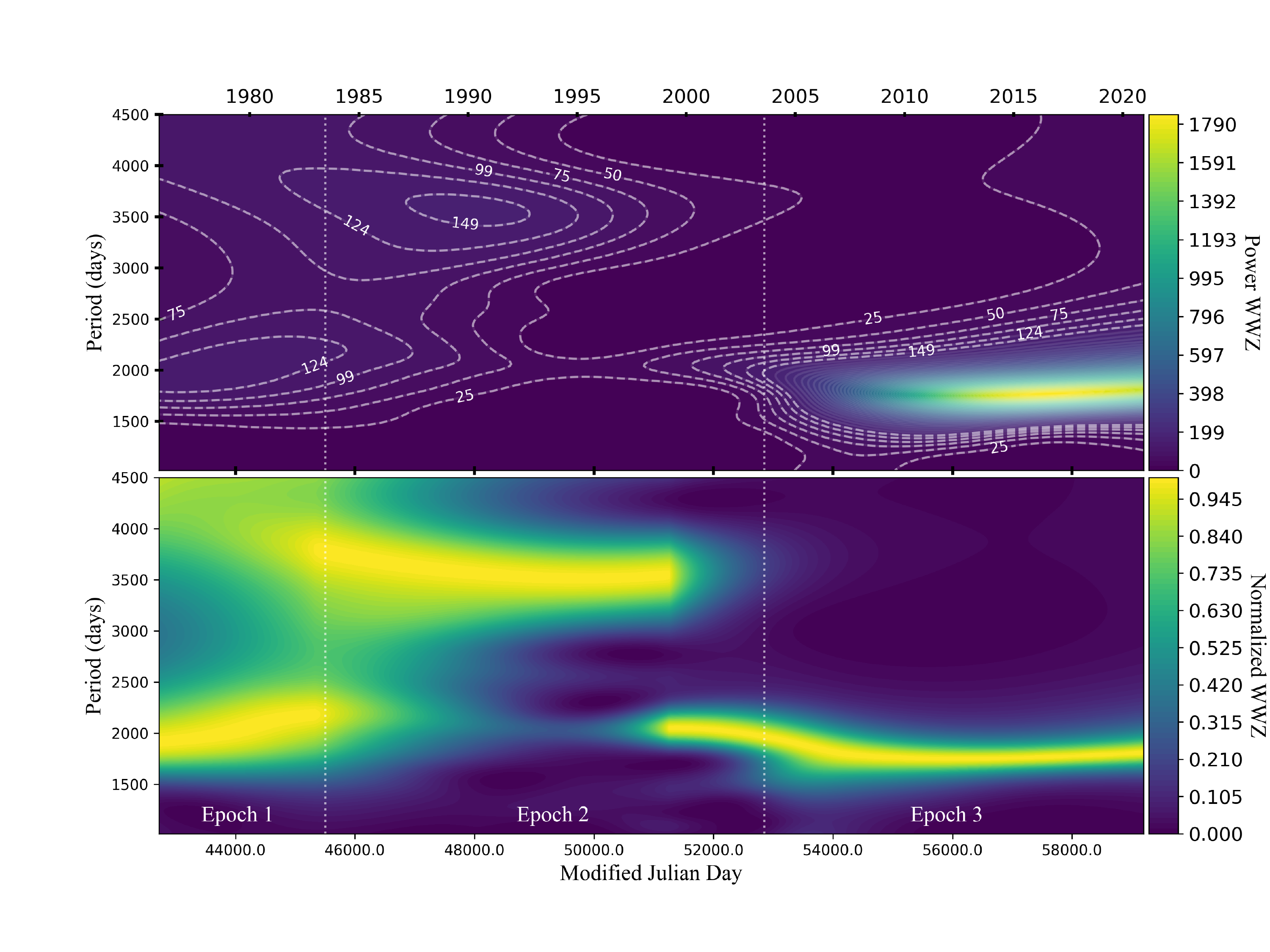}
\caption{ Plots of the WWZ statistic computed from the PKS~2131$-$021 45.1-year light curve. Upper Panel:  WWZ statistic plot, in which the color scale at the right and the contour levels on the plot represent the value of the WWZ statistic (see Appendix B).  In this panel all the features apart from the feature in epoch~3 are almost invisible  because they are more than an order of magnitude weaker than the epoch~3 feature.    Lower Panel:  Time-normalized WWZ plot of the 45.1-year light curve of PKS~2131$-$021.   Here we normalized each of the 164 100-day time bins by dividing the  power across all the periods in each 100-day time bin by the maximum power in the bin. Thus every time bin has a maximum power of unity.  This enables us to see clearly how the periods change over the whole 45.1-year span of the observations. }
\label{plt:wwz}
\end{figure}

\section{Analysis of the 14.5 GHz--15.5 GHz Radio Light Curve}\label{sec:analysis}

We have long been  aware of the dangers of over-interpretation of AGN light curves. Before analyzing our radio light curve of  PKS~2131$-$021, we outline the approach we have adopted to avoid the red noise pitfall, and other systematic issues that could undermine our key results.   

In order to make statistically robust estimates of the chance probability of occurrence of any feature in a radio light curve we need a process of simulating the light curve of any AGN such that all of the statistical and variational characteristics of the AGN are preserved in the simulations, while at the same time taking into account that  the data are not equi-spaced due to weather and hardware problems. We can then simulate a large number of light curves for any AGN and  estimate the probability of chance occurrence of any feature we observe in the light curve of any particular AGN. More details are given in Appendix \ref{app:significance}.

The problem of generating simulated light curves has been addressed in a number of papers  \citep{1999PASP..111.1347W,2002MNRAS.332..231U,2010MNRAS.404..931E,2014MNRAS.445..437M}, which simulate  light curves with PSD  of the same variability power law slope as observed in the AGN.  However the underlying PDF is Gaussian, and does not produce realistic light curves. Radio AGN light curves exhibit ``burst''-like events, that can often yield long-tailed PDFs.   

This problem was elegantly solved by \cite{2013MNRAS.433..907E} using a simple method that precisely reproduces light curves that match both the PSD and the PDF of the observed light curves. As desired, the final artificial light curves have all the statistical and variability properties of the observed light curves, and are statistically (and visually) indistinguishable from the true light curves. We have implemented this approach\footnote{\url{https://github.com/skiehl/lcsim}} and applied it to the light curve of PKS~2131$-$021 in our analysis.
For each epoch~we simulated light curves that have the same PSD, PDF, sampling, and ``observational'' noise as the real data of that specific epoch.
We are therefore confident that the significances we calculate are robust, having correctly taken into account the red noise in the observed PDF. 

In our analysis we have carried out a a detailed WWZ analysis \citep{1996AJ....112.1709F},  as well as a detailed GLS periodogram analysis \citep{1976Ap&SS..39..447L,1982ApJ...263..835S,2009A&A...496..577Z}, and we find the results to be fully consistent with our results derived from least-square sine wave fitting of the light curve.

\begin{figure}[!t]
\centering
\includegraphics[width=1.0\linewidth]{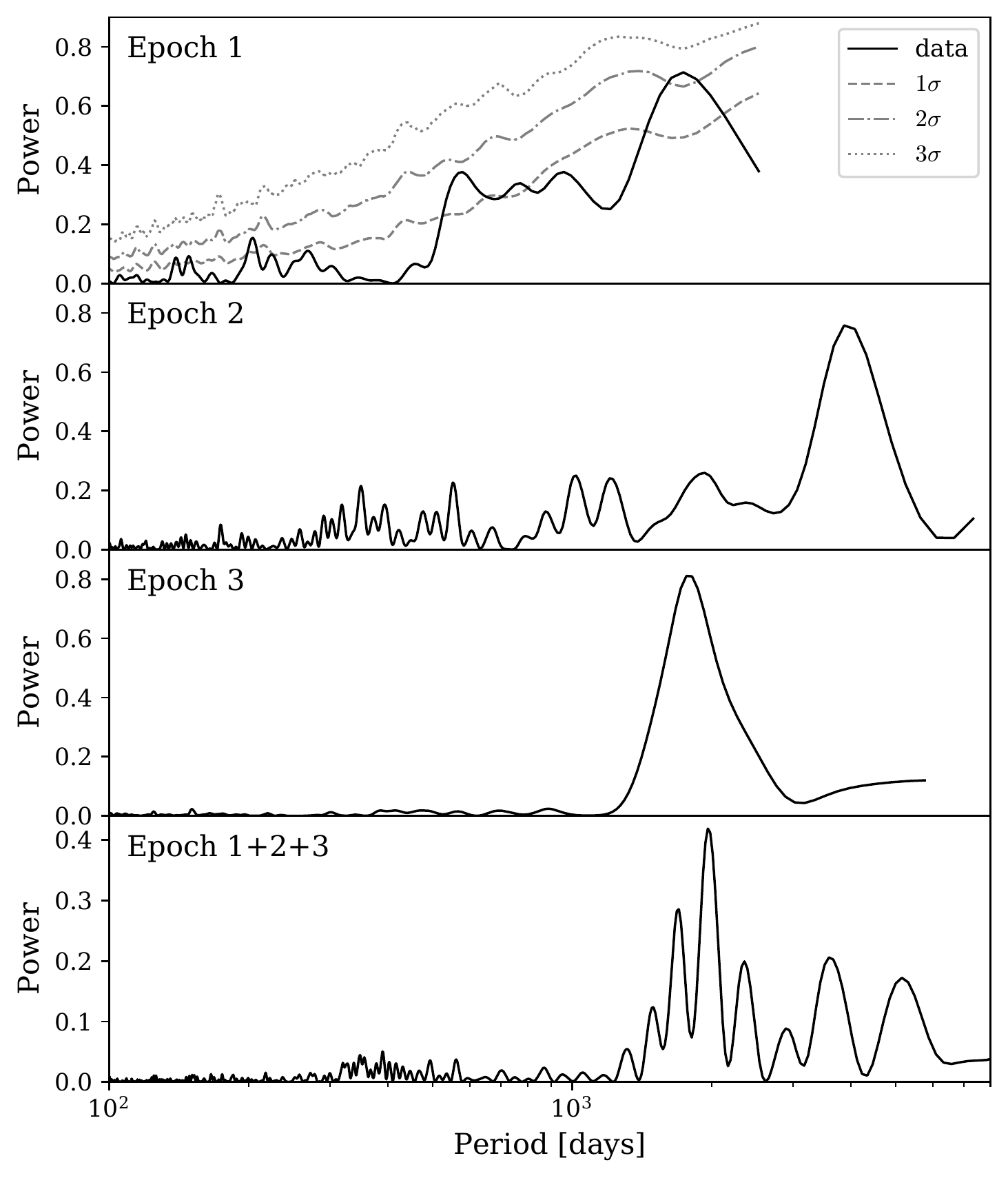}
\caption{Generalized Lomb-Scargle analyses of the light curve of PKS~2131$-$021. In the upper three panels we show the GLS analyses for the three distinct epochs of activity.
As is common practice, in the top panel the contours purport to show the significance levels of the features. The apparent significance of the peak in epoch~1 (top panel) is $2.3\sigma$, whereas the true significance is $1.06\sigma$ (see Table~\ref{tab:elsewhere}).  We do not plot contours in the middle and bottom panels because such  contours are misleading and further exacerbate the problem of the mis-identification of random features as significant peaks  (see \S~\ref{sec:caution}). The bottom panel shows the GLS power plot for the full 45.1-year duration of the observations, i.e. for epochs 1, 2 and 3 combined.  (see text).}
\label{plt:LSpowers}
\end{figure}

\subsection{WWZ Analysis of the Light Curve}\label{sec:WWZ}

The WWZ is a Z-transform \citep{ragazzini1952analysis} that uses the technique of wavelet analysis to detect time-dependent periodicity in data \citep{1996AJ....112.1709F}. We have performed a thorough WWZ analysis\footnote{\url{https://github.com/skiehl/wwz}} of the light curves of the three epochs and all combinations thereof. These showed very clearly that there are three distinct epochs. This is most easily seen in the WWZ plot covering the whole 45.1-year light curve shown in Fig.~\ref{plt:wwz}.  The power across this period varies strongly and the raw WWZ plot (upper panel) does not have the dynamic range to enable us to see the key features very clearly.  -- more details are given in Appendix B.  For this reason, we calculated  
time-normalized WWZ estimates by dividing the WWZ power in each 100-day time bin by the maximum power in the time bin in order to bring out the details of any changes in period (lower panel).   This WWZ plot supports the conclusion we arrived at by visual inspection of the light curve that there are three distinct epochs of activity in PKS~2131$-$021 over the 45.1-year span of our observations, with a periodic oscillation of approximately the same period dominating in epochs 1 and 3, and a periodic oscillation of about twice that period dominating in epoch~2 (see Table~\ref{tab:lsvalues}).  It is interesting to note that the boundary between epochs 1 and 2 is at the point where the WWZ power shifts from one period to another, whereas the boundary between epochs 2 and 3 is later than the point where the WWZ power shifts from one period to another. We ascribe the dominant variability in epoch~2 to red noise.   It is possible that the transition from epoch~2 to epoch~3 occurred earlier, but we decided to retain the boundary we picked out by eye since the periodicity of epoch~3 is very clear from this point on and not nearly as clear before that, so that we suspect that all of the interval we demarcate as epoch~2 is dominated by red noise.  Furthermore, we tried shifting the boundary between epoch~2 and epoch~3 between MJD 51200 and MJD 53800, and found that it had little effect on the results, as shown in the notes to Table~\ref{tab:sine_fits}. Clearly, the precise choice of the boundary between epoch~2 and epoch~3 is immaterial for the purposes of this study. 

Since the results of the WWZ analysis were entirely consistent with those from our GLS analysis and least-squares sine wave analysis, we do not reproduce further details of our WWZ analysis here other than to give the periods associated with the maximum power level in our WWZ analysis for the three epochs (see the notes to Table~\ref{tab:sine_fits}).

\subsection{Lomb-Scargle  Analysis of the Light Curve}\label{sec:lombscargle}

% \begin{deluxetable}{l@{\hskip 8mm}cccccc}
%\tablecaption{Lomb-Scargle Peak Period and Power and Significances from Simulations}
%\tablehead{epoch~& $P_{\rm peak}$ & Power & $N_{\rm sim}^{\rm tot}$ & $N_{\rm sim}$ & p-value & %$\sigma$\\
%&(days)&(max=1)&&&&}
%\startdata
%1 & 1737 & 0.71 & 10000 &  108 & 0.0108 & 2.30\\
%2 & 3863 &0.76 &10000 &  27 & 0.0027 & 2.78\\
%3 & 1763& 0.81 &100000  &  0 & $<10^{-5}$ & $>4.26$\\
%\enddata
%\tablecomments{The peaks with greatest power in each of the three epochs. No other peaks were seen
%with power greater than 0.4.  We therefore take 0.5 as a safe lower limit or accepTable~LS power in %our analysis. $N_{\rm sim}$ is the number of simulations with equal or greater power than observed %at the peak period. For comparison the WWZ analysis yields peak power at periods of 1740~days, %3919~days, and 1779~days, for epochs 1, 2 and 3, respectively.}
%\label{tab:lsvalues}
%\end{deluxetable}

We carried out a GLS periodogram analysis \citep{1976Ap&SS..39..447L,1982ApJ...263..835S} using the Generalised Lomb-Scargle () periodogram\footnote{\url{https://github.com/sczesla/PyAstronomy}} \citep{2009A&A...496..577Z} on each epoch~of the light curve of PKS~2131$-$021, as well as all combinations thereof. In order to distinguish it from the period of the periodicity, $P$,  we use the symbol $\mathcal{P}$ to denote the GLS power.
{ Following  \citet{1982ApJ...263..835S}, we evaluate the GLS at frequencies between $f_{\rm low} = 1 / T$, where $T$ is the total time of the corresponding epoch, and $f_{\rm high} = N / (2T)$, corresponding to one over twice the average time interval between measurements with $N$ the number of data points. The selection of a pseudo-Nyquist frequency as the highest frequency does not have any effect on the results as all relevant peaks are found at much lower frequencies. We sample this frequency range uniformly in steps of $\Delta f = 1 / (\zeta T)$. For our analysis we conservatively chose $\zeta=10$, since with this value we can be confident that we have adequately sampled periods up to the duration of the observations. For comparison,  \citet{1982ApJ...263..835S} used $\zeta=5$.}
{ 
We also tested the effect of the frequency grid spacing with  coarser and a finer grids, $\zeta=5$ and $\zeta=20$, and found only slight differences compared to the values for $\zeta=10$.  The variations with $\zeta$ were far too small to affect our conclusions in the analysis presented here.}

The results of the GLS analyses of epochs 1, 2 and 3, both independently and combined,  are shown in Fig.~\ref{plt:LSpowers}, and Table~\ref{tab:lsvalues}. There are about 1.5 cycles of the periodic variation in the Haystack+UMRAO data and about 3 in the UMRAO+OVRO data.   While this is a small number of cycles, our simulation procedure takes long-term fluctuations into account via the red tail in the spectrum, and so we are confident that our significance levels are robust.     Clearly, epoch~3 has  the statistically strongest periodicity, and the similarity of the period with that of epoch~1, as well as the phase correspondence, which are what immediately caught our attention when we first saw the Haystack observations, when taken together, are highly significant, as we will show.

In the combined GLS analysis if the bottom panel of Fig.~\ref{plt:LSpowers}  the largest peak corresponds to the epoch~3 peak, the second largest peak corresponds to the epoch~1 peak and the third largest peak corresponds to the epoch~2 peak. All of these three peaks in the combined analysis of the bottom panel are slightly shifted in period relative to their period values in the individual epoch~GLS analyses.
Given that the epoch~3 peak is the only peak detected at greater than 3$\sigma$ significance, and that the spectrum has a powerful red noise component, the peaks other than the epoch~3 peak, should not be regarded as physically significant unless supported by other data, as is true for the epoch~1 peak, but not for any of the other peaks seen in the combined analysis. This, plus the WWZ analysis presented in Fig.~\ref{plt:wwz}, illustrates the potential danger of carrying out GLS, or WWZ, analyses on astronomical light curves without adequately simulating the variability in the object and, in addition, applying rigorous statistical criteria.

\subsection{Sine-Wave Least-Squares Fitting of the Light Curve}\label{sec:sinewave}

We fitted the light curves in the three epochs with a sine wave function:
\begin{equation}
S_{{\rm model}, i} = A \sin(\phi_i-\phi_0) + S_0,
\end{equation}
where $\phi_i = 2\pi(t_i-t_0)/P$ is the phase of the $i$th data point.
Our model has five free parameters: $P$ -- period, $A$ -- amplitude, $\phi_0$ -- phase of the sine wave, $S_0$ -- mean flux, and $\sigma_0$ -- characteristic amplitude of an intrinsic AGN variability. We keep $t_0=51000$ fixed (which is in the midpoint between the start of epoch~1 and end of epoch~3 observations). We find the best-fitting parameters by maximizing the following likelihood function:
\begin{equation}
\ln\mathcal{L} = -\frac{1}{2}\sum_{i=1}^{N} \left(\frac{(S_i-S_{{\rm model}, i})^2}{\sigma^2_i+\sigma_0^2} + \ln{(\sigma^2_i + \sigma_0^2)}\right).
\end{equation}
The uncertainties are estimated using the \textsc{emcee} sampler \citep{Foreman2013} and represent the 68\% confidence range of the marginalized posterior distribution.  
The $\sigma_i$ are taken from the observations, assuming independent Gaussian errors, and the intrinsic variability is represented by Gaussian white noise of variance $\sigma_0^2$.    We have not used the red noise spectrum because to include it in the fit would mean including the full covariance matrix rather than just the diagonal term. We know that including a red noise term of the type seen in epoch~2 can change the period at the $\sim$ 10\% level over periods of a few years (see \S~\ref{sec:varper}).

Fit results, separately for epochs~1, 2 and~3, are presented in Table~\ref{tab:sine_fits}. Note that the uncertainties of $\phi_0$ are relatively large because there are strong degeneracies between period and phase, especially  as the adopted $t_0$ is relatively far from the actual observations. If we chose $t_0$ to be closer to epoch~1 (epoch~3) observations, the respective uncertainties of $\phi_0$ would be much smaller. The degeneracy between the period and phase reflects the fact that, over time, information about the phase is lost.

We also simultaneously fitted both epoch~1 and epoch~3 data by fitting a single wave, with the same period and phase extending across both epochs, but permitting the amplitude and offset between epoch~1 and epoch~3 to vary.    Results of this joint fit are presented in the last column of Table~\ref{tab:sine_fits}.

\begin{table*}[!t]
\centering
\caption{Sine-fitting results for epoch~1, epoch~2, epoch~3, and the joint epoch~1 + epoch~3 data sets.}
\begin{tabular}{lrrrr}
\hline \hline
& epoch~1 & epoch~2 & epoch~3 & epoch~1 + epoch~3\\
\hline
$P$ (days)           & $1729.1 \pm 32.4$ & $3779.1 \pm 46.0$ & $1760.4 \pm 5.3$  & $1737.9 \pm 2.6$  \\
$\phi_0$             & $0.89 \pm 0.46$   & $0.35 \pm 0.08$   & $0.60 \pm 0.07$   & $0.88 \pm 0.03$   \\
$A$ (epoch~1)        & $0.709 \pm 0.047$ & \dots             & \dots             & $0.679 \pm 0.045$ \\
$S_0$ (epoch~1)      & $2.553 \pm 0.036$ & \dots             & \dots             & $2.584 \pm 0.034$ \\
$\sigma_0$ (epoch~1) & $0.333 \pm 0.022$ & \dots             & \dots             & $0.337 \pm 0.023$\\
$A$ (epoch~2)        & \dots             & $0.392 \pm 0.020$ & \dots             & \dots \\
$S_0$ (epoch~2)      & \dots             & $1.724 \pm 0.021$ & \dots             & \dots \\
$\sigma_0$ (epoch~2) & \dots             & $0.140 \pm 0.009$ & \dots             & \dots\\
$A$ (epoch~3)        & \dots             & \dots             & $0.400 \pm 0.007$ & $0.400 \pm 0.007$ \\
$S_0$ (epoch~3)      & \dots             & \dots             & $2.225 \pm 0.005$ & $2.229 \pm 0.005$ \\ 
$\sigma_0$ (epoch~3) & \dots             & \dots             & $0.118 \pm 0.004$ & $0.120 \pm 0.004$ \\
\hline
\end{tabular}
\tablecomments{ We also determined the least squares sine fit to epoch~3 for shifted boundaries between epoch~2 and epoch~3 at MJD~51200 and MJD~53800, with the results $P=1762.9 \pm 6.1$d and $P=1756.0 \pm 5.5$d, respectively. These may be compared with the result above of $P=1760.4 \pm 5.3$d for the boundary set at MJD~52850.  The peak periods identified in our WWZ analyses of epochs 1, 2,  and 3 were 1740  days, 3919  days, and   1779  days, respectively. Those in our GLS analyses were 1730  days, 3937   days, and 1788  days, respectively.}
\label{tab:sine_fits}
\end{table*}

\begin{deluxetable*}{c@{\hskip 8mm}ccccccc}
\tablecaption{Probabilities and Significance Levels of GLS Tests Computed from Simulations with Matched Red Noise Tail}
\tablehead{Test&Test& GLS $\mathcal{P}_{\rm peak}$ & Period Range ($\Delta P$)& Total& Number of simulations& p-value & Significance\\
Number&&(max=1)&(days)&Simulations&that pass test&&($\sigma$)}
\startdata
1.1& epoch~1 $\mathcal{P}_{\rm peak}$, $p_{\rm sim} \leq p_{\rm peak}$ & 0.71&All&10000  &  1446 & $1.45 \times 10^{-1}$ & 1.06\\
1.2 & epoch~2 $\mathcal{P}_{\rm peak}$, $p_{\rm sim} \leq p_{\rm peak}$ &  0.76 &All&10000  &  632 & $6.32 \times 10^{-2}$ & 1.53\\
1.3& epoch~3 $\mathcal{P}_{\rm peak}$, $p_{\rm sim} \leq p_{\rm peak}$ &  0.81 &All&100000  &  40 & $4.0 \times 10^{-4}$ & 3.35\\
2&epoch~1 $\mathcal{P}_{\rm lim}$, $\Delta P_{\rm epoch~3}$ &$\ge 0.50$ &1661.9 -- 1858.9&10000  &  197 & $1.97 \times 10^{-2}$ & 2.06\\
3& 1.3+2& -&-&-&-&$7.88 \times 10^{-6}$&4.32\\
4& 3+phase& -&-&-&-&$1.58 \times 10^{-6}$&$4.66$\\
\enddata
\tablecomments{  $\mathcal{P}$ is the GLS power and $\Delta P$ is the range of periods included in the test. In Tests 1.1, 1.2, and 1.3 we count simulations at all periods with p-values less than the p-value of the peak in that epoch.  Since the significance of the peak in epoch~3 was so high, we needed 100,000 simulations to determine the significance. The low individual significance levels for epochs 1 and 2, shown in Tests 1.1 and 1.2, indicate that they are easily produced by red noise.  Test 2 is the number of simulations of epoch~1 that were seen in the $\pm 3\sigma$ wide  period window centered on the epoch~3 period ($P=1760.4$~days).}
\label{tab:lsvalues}
\end{deluxetable*}

Prior to MJD 45500 (epoch~1) the strong sinusoidal variation seen by Haystack+UMRAO has period $P_1=(1729.1\pm 32.4)$~days; in epoch~2 (45500$<$MJD$<$52850), the UMRAO data show sinusoidal variability, but no periodicity at the frequency or in phase with that seen in the first epoch~and third epochs. We therefore  think that this is  a red noise phenomenon, and we will show that it fails our $3\sigma$ criterion by a substantial margin; in epoch~3  (MJD$>$52850)  strong sinusoidal variation returns with period $P_3=(1760.4\pm 5.3)$~days. In addition, the phase of the sinusoidal variation from epoch~3, when extrapolated back to epoch~1, matches to within 10\% of $P$, or 20\% of $P/2$, which is the relevant number here since we associate the extrapolated curve with the nearest peak.  Combining the Haystack, UMRAO, and OVRO data from epoch~1 and epoch~3 yields a period $P_{13}=(1737.9\pm 2.6)$~days, i.e. the fractional uncertainty in period $\delta P/P \sim 1.5 \times 10^{-3}$.

\subsection{The Significances of the Observed Periodicities}\label{sec:sign1}

To calculate the significance of the power levels observed in epochs 1, 2, and 3, we count the number of simulations in which the p-value, $p_{\rm sim}$, of the strongest peak in the simulation is less than  the observed peak p-value, $p_{\rm peak}$,  for that epoch. The procedure is explained in more detail in Appendix~\ref{app:significance}. The results are shown in Table~\ref{tab:lsvalues}. We see that the peaks in epoch~1 and epoch~2  are  significant only at the $1.06\sigma$ and $1.53\sigma$ levels, respectively (Tests 1.1 and 1.2). However, the peak in epoch~3  is significant at the $3.35\sigma$ level (Test 1.3). This makes it clear that PKS~2131$-$021 has a periodicity in epoch~3 that is not due to red noise, which  must therefore be associated with some physical mechanism in the object.

We next turn to the good agreement in the periods, straddling the 20-year gap in periodic fluctuations, seen in epochs 1 and 3 (see Table~\ref{tab:sine_fits}), which, to our knowledge, is  unprecedented in observations of AGN.
The question we  address is this: having observed the periodicity in epoch~3, which is what first drew our attention to this source,  what is the probability of observing  the same periodicity in epoch~1, at an acceptable~LS power level and within  the $3\sigma$ period window of our epoch~3 period?

We first determine what we deem to be an acceptable~GLS power level, $\mathcal{P}_{\rm lim}$. From Table~\ref{tab:lsvalues}, we see that, of the three periodicities we have been discussing,  the lowest power level  is that of the peak in epoch~1, at $\mathcal{P}_{\rm peak} = 0.71$. From Fig.~\ref{plt:LSpowers} we see that, apart from the largest peak in each epoch, all the other peaks in power are below the $\mathcal{P} = 0.4$ power level.  We therefore select a threshold level of power of $\mathcal{P}_{\rm lim} = 0.5$ as a power level suitably above the level of sidelobes in  the GLS spectrum and suitably low that we do not miss any strong sinusoidal features in the light curves.  We will therefore count all simulations that give an GLS power level $\mathcal{P_{\rm lim}}\ge 0.5$ in epoch~1, in our test of the significance of the epoch~1 periodicity, given the epoch~3 periodicity.%(i.e. in Test 2 in Table~\ref{tab:lsvalues}).

We now turn to the range of periods to be considered. From Table~\ref{tab:sine_fits} we see that the uncertainty in the difference between $P_1$ and $P_3$ is $\sqrt{32.4^2+5.3^2}=32.8$~days.  Thus the $3\sigma$ uncertainty in the difference is 98.5~days.  
Since we wish to calculate the probability of observing the periodicity in epoch~1, given that it had been observed in epoch~3, we will count simulations in epoch~1 where the strongest peak lies in the period range $P_3\pm 98.5$~days, i.e. from 1661.9~days to 1858.9~days.  So our two criteria are GLS power $\mathcal{P}\ge \mathcal{P}_{\rm lim} = 0.5$ and 1661.9~days $< P <$ 1858.9~days.

The probability of observing in epoch~1 a peak within the $\pm 3\sigma$ window of the periodicity observed in  epoch~3 and with $\mathcal{P} \ge 0.5$ is shown in Test 2 of Table~\ref{tab:lsvalues}  to be $ 1.97 \times 10 ^{-2}$, which is significant at the $2.06\sigma$ level. We may now combine Test~1.3 and Test~2 to determine the probability of observing both the observed periodicity in epoch~3 and a periodicity in epoch~1 with period within the $\pm 3\sigma$ window of the periodicity found in epoch~3. This is determined in Test~3 by multiplying $p(\text{Test}~1.3) \times p(\text{Test}~2) = 7.88 \times 10 ^{-6}$, which is significant at the $4.3\sigma$ level.
When we add the correspondence in phase to within one-fifth of a half-cycle, the probability drops to $0.2 \times p(\text{Test}~3) = 1.58 \times 10^{-6}$, which is significant at the $4.6 \sigma $ level. 

We believe that these significance estimates, based as they are on simulations that have the same PSD and the same PDF as the observed light curves in epochs 1, 2 and 3, and therefore with the same red noise  tail as PKS~2131$-$021, are robust, and therefore that this analysis demonstrates conclusively  that the periodicities observed in epochs 1 and 3 are connected and that there is a physical process that maintains this period over our 45.1-year observing period even when it is not manifested in the light curve.

From Table~\ref{tab:sine_fits} we see that the observed difference in period between the epoch~3 data and the epoch~1 + epoch~3 data is significant at the $3.8\sigma$ level, so this might indicate that the period is slowly changing. Separate sine wave fits to the first half and the second half of the OVRO data show a difference of $\sim$10\% in period. PKS~2131$-$021 is a blazar, and, as such, highly variable even in the absence of the periodicity we have been discussing, as is easily seen from the deviations from sinusoidal variations observe in epochs 1 and 3 and throughout epoch~2. Thus we ascribe the apparent changes in period to random red noise variability in this blazar, such as that seen in epoch~2. {\it Indeed,  because of the strong red noise component in their variability,   such variations in the observed period in radio light curves of an SMBHB in a blazar like PKS~2131$-$021 are to be expected,  but these should average out in the long term.}

The reader may have noticed the apparent harmonic relationship between the epoch~2 and epoch~1+3 periodicities. In this paper we do not wish to draw undue attention to this for two reasons:
\vskip 2pt
\noindent
1. The epoch~2 periodicity has significance of only $1.53\sigma$, as shown in Table~\ref{tab:lsvalues}.
We do not discuss this periodicity since we think any discussion runs counter to the approach we are adopting in this paper.  We  encourage workers in this field to carefully assess red noise before entering discussions of apparent periodicities.
\vskip 2pt
\noindent
2. There is no obvious phase relationship in between the real, highly significant, periodic signals we see in epochs 1 and 3 and the low-significance apparent periodicity in epoch~2.

{\it We re-emphasize here our deliberate decision to consider only signals that are demonstrably not due to red noise, i.e. only those having significance greater than $3\sigma$, unless supported by other evidence.}

\begin{deluxetable}{l@{\hskip 8mm}cccc}
\tablecaption{Single Period (spurious)  and All Period (true) Probabilities}
\tablehead{ epoch~and test&$N_{\rm tot}$&$n_{\rm pass}$&p-value&$\sigma$\\
}
\startdata
epoch~1 single period &10000 &108&$1.08 \times 10^{-2}$&2.3 \\ 
epoch~1 all periods &10000 &1446&$1.45 \times 10^{-1}$&$1.06^\dagger$ \\ 
epoch~2 single period &10000 &122&$2.20 \times 10^{-3}$&2.85 \\ 
epoch~2 all periods &10000 &632&$6.32 \times 10^{-2}$&1.53$^\dagger$ \\ 
epoch~3 single period &100000 &0&$< 10^{-5}$&$>4.26$ \\ 
epoch~3 all periods &100000 &40&$4.00 \times 10^{-4}$&$3.35^\dagger$ \\ 
\enddata
\tablecomments{The tests using all periods are the ``Look Elsewhere'' tests.$^\dagger$ these are the true significances, the others are totally spurious unless the periods have been selected ``a priori''.}
\label{tab:elsewhere}
\end{deluxetable}

\subsubsection{The ``Look Elsewhere'' Effect}\label{sec:caution}
 In the GLS analysis the periods are quantized because they are based on discrete frequencies (see Appendix D of \cite{1982ApJ...263..835S}).  It is  common, in studies of QPOs, to see significance contours, derived from simulations, plotted on GLS and WWZ plots.    To the best of our knowledge, such contours are determined by counting the number of simulations at each quantized period and dividing by the total number of simulations, as we have done in Fig.~\ref{plt:LSpowers} (top panel) for the sole purpose of illustrating this common pitfall.  This is a legitimate  approach   {\it only} if that periodicity has been pre-selected, for example a known period in a double star,  with known uncertainty in the period, in which case the value of $\zeta$ must  be chosen so as to include the corresponding range of periods at the peak period sampling interval (see \S~\ref{sec:lombscargle}). However,  there are two problems with this approach in general. The first is that since the sampling in periodicity is discrete and depends on $\zeta$ the single-period p-value $\propto 1/\zeta$; and the second is that   if there is no {\it a priori} reason for selecting a particular period, then the significance of any peak  must take into account all of the simulations in which the largest peaks have p-values that are less than or equal to the p-value of the peak under consideration, as explained in the Appendix. 
Our three epochs on PKS~2131$-$021 provide a striking example of the necessity for taking the p-values of all the simulated light curves into account when assessing the significance of an observed feature. The results are shown in Table~\ref{tab:elsewhere}. The true significances address the question ``what is the probability of finding a peak of {\it any} periodicity  that has a p-value less than or equal to that of the observed peak?'',  in other words on has to look elsewhere than just at the period corresponding to the peak in the GLS plot. The problem of using single period probabilities, and of propagating these by plotting misleading significance curves on GLS plots,  further exacerbates, in addition to the red noise problem, the problem of random noise peaks being considered as if  they are physically significant.

\begin{figure*}[ht!]
\centering
\includegraphics[width=0.8\linewidth]{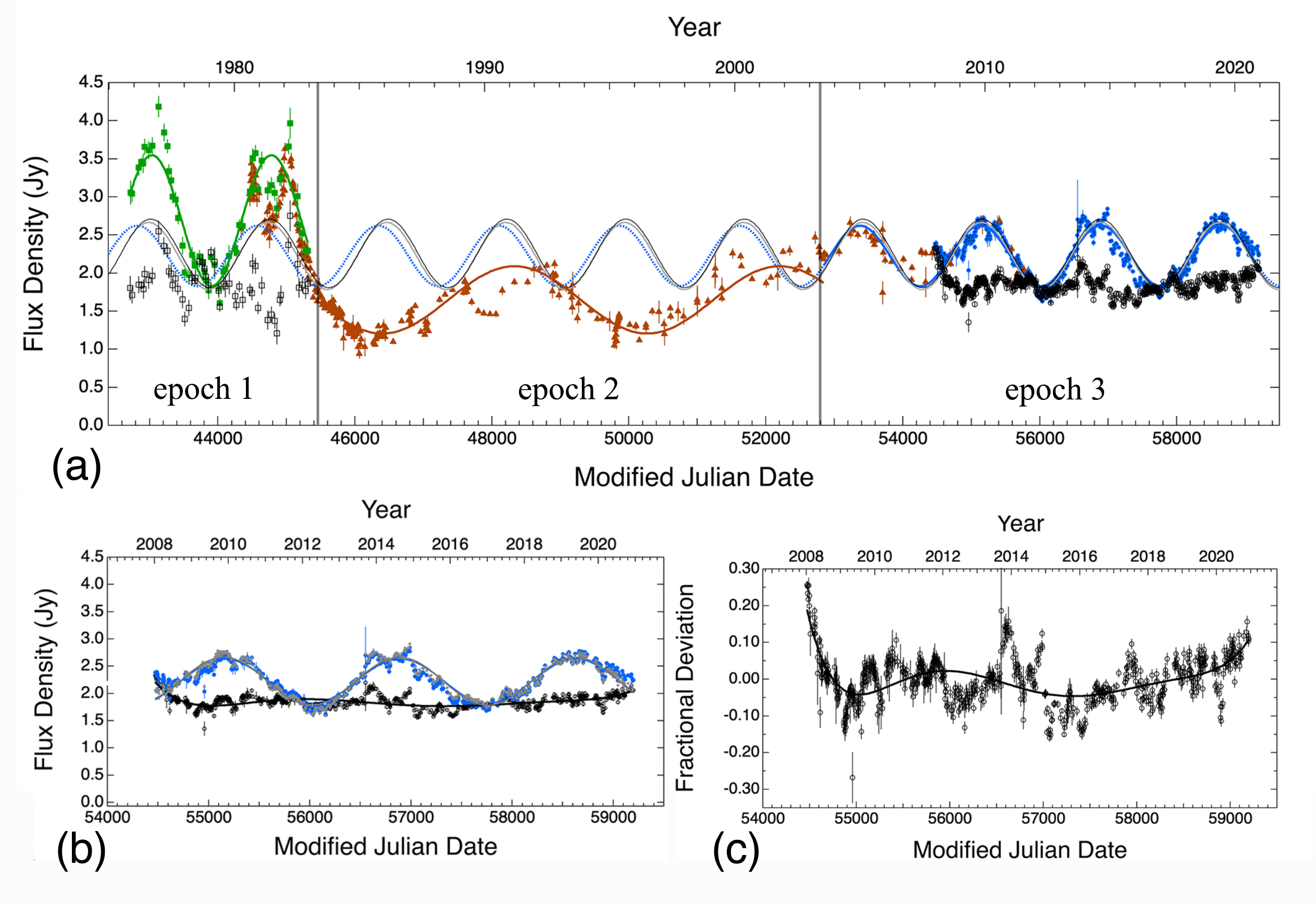}
\caption{Period variations in the 14.5~GHz - 15.5~GHz light curve of PKS~2131$-$021. (a) The Haystack data are shown by the green squares, the UMRAO data by the brown triangles and the OVRO data by the blue dots.   The solid green line shows the least-squares sine wave fit to the Haystack data, and the dotted blue line shows the least-squares sine wave fit to the OVRO data  extrapolated backwards to provide a phase comparison with the Haystack data.   In epoch~2  we show a 6-th degree polynomial fit to the light curve, and it is seen to be sinusoidal in character. This oscillation has about twice the period of that in epoch~1, but it  bears no obvious phase relationship with the oscillation of epoch~1, and it also bears no obvious phase relation with the epoch~3 periodicity. In epoch~3 we have fitted the combined Haystack and OVRO data (black sine wave). This fit, adjusted for the different amplitudes of the Haystack and OVRO periodic signals, has been  subtracted from the OVRO data to provide the residual light curve in absence of the periodic signal (black data points). The  black curve in (b) and (c) shows a 6-degree polynomial fit to the residual OVRO data. The gray curve is the sine wave fit to the OVRO data after correcting for the slowly varying component in the residual signal of epoch~3 (black point), which brings the phase  into alignment with that of epoch~1, {\it as expected}. (b) the blue points and blue curve show the raw OVRO data and sine wave fit, and the gray points show the corrected OVRO light curve after applying the slowly varying component given by the black 6-degree polynomial fit to the residuals. The gray sine wave is the least-squares fit to the adjusted (gray) OVRO data. (c) Fractional flux density deviation from the mean value with the OVRO sine wave subtracted. This shows the fractional flux density variations that would be required to bring the Haystack and OVRO data into phase coherence.}
\label{plt:light curves 3}
\end{figure*}

\subsection{Variations in the Period Due to Noise}\label{sec:varper}

The fits of the sine waves shown in Fig.~\ref{plt:light curves}, while good, are not perfect.
There is a clear epoch, from 2015  to 2017 when the sine wave is systematically above the data, and similarly in 2020 the data are systematically above the sine wave. If the two halves of the OVRO data are analyzed separately the periods differ by $\sim 10\%$. Does this mean that PKS~2131$-$021 is simply another QPO? The unique properties we have drawn attention to above suggest it is not.  

We now provide an illustration,  based on the variability during epoch~2, that refines the period derived for epoch~3 to $P=(1737.6\pm 3.6)$~days, shown by the gray curve in Fig.~\ref{plt:light curves 3}(a) and (b), in good agreement with the period and phase observed over the whole 45.1 years. This is merely  an example of what must be  happening, under our hypothesis of an SMBHB. We go through the following argument to illustrate the effect that we think the non-periodic varying components must be having on the observations. Of course, this does not prove that PKS~2131$-$021 is an SMBHB. But the exercise is illuminating, and in our view does therefore strengthen the case for an SMBHB on the grounds of plausibility. 

We have fitted the period when the sinusoidal variations we are investigating were absent using the epoch~2 data between MJD 45500 (1983 June 15) and MJD 52850 (2003 July 30), and found that a 6-degree polynomial, shown by the brown curve in Fig.~\ref{plt:light curves 3}(a),  follows the slowly varying component well, whereas lower degree polynomials do not. We adopted the least-squares sine wave fit to epoch~1 + epoch~3 and subtracted this, adjusted for the amplitudes of the Haystack and OVRO periodic signals, from the OVRO data to give the residual light curve shown by the black points in Fig.~\ref{plt:light curves 3}(a).  This shows the flux density variations in PKS~2131$-$021, under the SMBHB hypothesis and in the absence of the sinusoidal signal. We then adopted the same approach with the OVRO data, as with the UMRAO data, and  fitted a 6-degree polynomial to the residual light curve  to give the black curves shown in Fig.~\ref{plt:light curves 3}(b) and~(c).  These show what the slowly varying components of the source were doing, under the SMBHB hypothesis and apart from the periodic signal. Correcting for the slowly-varying components we derive the gray points shown in Fig.~\ref{plt:light curves 3}(c).  The sine wave fit to the corrected OVRO data are shown by the gray curve in Figs. \ref{plt:light curves 3}(a) and (b), which has period $P=1737.6\pm 3.6$~days.  Thus, {\it as expected}, it has a period closer to the period of the epoch~1 + epoch~3 sine wave fit ($P=1737.9\pm 2.6$~days) than the original sine wave fit to the epoch~3 data ($P=1760.4\pm 5.3$~days). In addition, and  {\it as expected}, the phase during epoch~1, as shown by the gray curve in Fig.~\ref{plt:light curves 3}(a) is much closer to that of the joint epoch~1 + epoch~3 fit (black curve) than the original epoch~3 fit (blue dotted curve). This example simply shows how small slow variations in the non-periodic signal can have a significant effect on the period and phase of the sinusoidal variations in the periodic signal in PKS~2131$-$021, a point which, while obvious, is worthwhile investigating to illustrate the levels of correction that are needed to achieve much better coherence.  The fractional changes in the flux density required for
perfect coherence, which would be grossly over-fitting the data, are shown  by the black points in Fig.~\ref{plt:light curves 3}(c). Note that they are nearly all $<10\%$.

\vskip 2pt
\noindent
Itemization of the above procedure:\footnote{This helpful summary was provided by our anonymous referee}
\vskip 2pt
\noindent
1. We fitted separate least-squares sine waves to the epoch~1 and epoch~3 data.
\vskip 2pt
\noindent
2. We subtracted these sine waves from the Haystack data of epoch~1 and the OVRO data of epoch~3.
\vskip 2pt
\noindent
3. We assumed that the sinusoid represented all of the SMBHB emission, and therefore that the SMBHB signal was now removed from the Haystack data and the OVRO data.
\vskip 2pt
\noindent
4. We assume epoch~2 is already devoid of SMBHB signal. Thus in Fig.~\ref{plt:light curves 3} the black data in epoch~1, the brown data in epoch~2, and the black data in epoch~3, represent the data with the SMBHB sinusoid subtracted.
\vskip 2pt
\noindent
5. We fitted a 6-order polynomial to the black data in epoch~3 in Fig.~\ref{plt:light curves 3}. The 6-order polynomial is thus assumed to represent the red noise variation of the SMBHB.
\vskip 2pt
\noindent
6. We then subtracted the polynomial fit given by the black line from the original data, which gives the grey points.
\vskip 2pt
\noindent
7. Finally, we did a least squares sine wave fit to obtain the period for the grey points. 

\vskip 2pt
\noindent
{\it Summary:} Our motivation in carrying out the above procedure is as follows. We have, quite deliberately, ``made the data fit the SMBHB model'' in order to provide an illustration of what we think the non-sinusoidal random flux density variations very likely were during the OVRO observations, and then to compare these to the non-sinusoidal random variation during epoch~2. This does not prove that PKS~2131$-$021 is an SMBHB, but, given the significance of the similarities in period and phase in epochs 1 and 3, as shown in Table~\ref{tab:lsvalues}, this is almost certainly the case.  Thus, understanding and anticipating the effects of the non-sinusoidal variability in a jetted-SMBHB is bound to be important as this field opens up.

It is interesting to note that the periodic fluctuations occur at the times of highest flux density in the overall light curve, so that it appears that the periodic signal is in addition to the usual flux density level in this source. Thus on our SMBHB model we are proposing that PKS~2131$-$021 has an extremely stable~period and that the small changes in period that we see are due to other sources of variation that distort the pure sine wave variation.  We also propose that the emission associated with the periodic variability can turn off and on in a time that is short compared to the period, and that the amplitude of the periodic signal can vary significantly between epochs of periodic emission.

Our observations thus lead to three strong predictions based on the SMBHB hypothesis: (1) we expect the future long-term averaged period to agree with our measured period spanning 45.1 years of observations ($P=1737.9\pm 2.6$~days); (2) in the short term we expect to see small ($\sim 10\% $) variations in the period due to the corrupting effects of the other varying emission components in this object;  and (3) we expect the SMBHB sinusoidal signal to continue to appear and disappear at random times.

\subsection{The Long-Term Stability of the Periodic Variations in the Light Curve}\label{sec:stability}

It is clearly important to investigate the stability of the periodic variations seen in PKS~2131$-$021, since this has potential implications for the SMBHB hypothesis,   under which the period should decrease with time.    For the purposes of comparison we will denote the periods determined in our least-squares sine fitting analysis in the different epochs (see Table~\ref{tab:sine_fits}) by $P_1 = 1729.1\pm 32.4$~days, $P_2 = 3779.1\pm 46.0$~days, $P_3 = 1760.4\pm 5.3$~days, and $P_{13} = 1737.9\pm 2.6$~days.

The most interesting of these is $P_{13}$ since it represents a single least-squares sine wave fit across the 45.1-year span.  On the hypothesis that PKS~2131$-$021 is an SMBHB, this represents our best estimate of the observed orbital period. On the SMBHB hypothesis we therefore predict that, in future long-term monitoring, this period will be maintained until the effects of gravitational radiation produce a noticeable reduction in the period. 

We therefore predict on the SMBHB hypothesis, that the next minimum is occurring now (October 2021), and we will be able to confirm this in the next 6 months. The  next maximum is predicted in February 2024, and the following minimum in July 2026. It should be possible to confirm or disprove the SMBHB hypothesis within the next 5-10 years, provided that we do not go back into an epoch~2-like phase in which this periodicity vanishes.

We now consider the stability of the period as can best be determined from the observations.
We have already seen that ~$\sim 10\%$ variations occur over time spans of a few years, which we ascribe to the corrupting effects of the variability in PKS~2131$-$021 that are unrelated to the underlying periodicity we seek to study.  Taking their associated uncertainties into account, we see that $\delta P_{13} = P_3-P_1 = 31.3 \pm 32.8$~days.  So the $3\sigma$
window is  $\delta P_{13}= -67.2 \rightarrow +129.8$~days.

On the SMBHB hypothesis we are interested in negative $\delta P_{13}$, and in this case  our  fractional $3\sigma$ limit  is $\delta P_{13}/P_{13} = -67.2/1737.9= -3.9 \times 10^{-2}$. The midpoint of epoch~1 is MJD=44109.8, and that of epoch~3 is MJD=56023.6, so the span between the midpoints at which $P_1$ and $P_3$ were measured is 11913.8~days, or $\delta t =32.6$ years.  Thus the $3\sigma$ upper limit on the fractional period decrease per year is $\delta P_{13}/P_{13}\delta t=-1.19 \times 10^{-3}$ per year.  This is the maximum rate of fractional period decrease per year consistent with our observations.  From \cite{maggiore2008}, using our rest-frame quantities, we derive a  chirp-mass upper limit of $5.4 \times 10^9 M_\odot$.  This may be compared with the slightly lower upper limit on the chirp mass derived from the gravitational wave limit discussed in \S \ref{sec:nanograv}.

\subsection{Summary of Our Approach and Statistical Tests}\label{sec:statsum}
In this section, before presenting our model for PKS~2131$-$021 and its implications,  we summarize briefly our approach and the logical thread of our statistical tests.
The key steps in our approach are:
\vskip 2pt
\noindent
1. The recognition of the role of red noise in the $\gamma-$ray, optical, and radio,  light curves of blazars, as exemplified in the work of \cite{2016MNRAS.461.3145V} and \cite{2019MNRAS.482.1270C}.  As a consequence we require strong statistical evidence that any apparent periodicity is not simply the result of red noise in the light curve, and we apply a $3\sigma$ threshold that must be met by any apparent periodicity before it should be discussed.  Without such a threshold, workers in this field will expend a great deal of time and energy seeking physical explanations for apparent periodicities that are the result of random fluctuations and have no other physical significance. Consideration of
such phenomena will obscure the physics and hinder progress in the field.
\vskip 2pt
\noindent
2. The recognition that there are three distinct epochs in the 45.1-year radio light curve of PKS~2131$-$021, as indicated by the periodicities seen in Fig.~\ref{plt:light curves}. The WWZ analysis of \S~\ref{sec:WWZ}, including Fig.~\ref{plt:wwz}, and the analysis of different MJD dates for the transition from epoch~2 to epoch~3 given in \S~\ref{sec:WWZ}, and in the Notes to Table~\ref{tab:sine_fits}, in addition to our original visual inspection, are our justification for this.
\vskip 2pt
\noindent
3. Starting with epoch~3, the epoch~that initially drew our attention to the periodicity in PKS~2131$-$021, we carried out a GLS analysis making no assumptions about periodicities, and considered all periodicities via the ``look elsewhere'' effect. We found that the probability at, or above, the observed power level in the GLS analysis is $4 \times 10^{-4}$, i.e. it is significant at the $3.35\sigma$ level, and thus it is above the threshold of point \#1 above.
\vskip 2pt
\noindent
4. We then investigated the sinusoidal periodicity of epoch~1 as follows: we selected a $\pm 3\sigma$-wide periodicity window, based on the uncertainties in the least squares periodicities in epoch~1 and 3, and determined the probability of a chance periodicity at or above the GLS power observed in epoch~1 in the window of this width and centered on the periodicity of epoch~3. This probability is $1.97 \times 10^{-2}$, i.e. it is significant at the $2.06 \sigma$ level.  
\vskip 2pt
\noindent
5. Since the data from epochs 3 and 1 are independent, the probability of the agreement in periods observed in epoch~1 and epoch~3 is given by the product of the two probabilities, and is $7.88 \times 10^{-6}$, i.e. it is significant at the $4.32 \sigma$ level.
\vskip 2pt
\noindent
6. Finally we took into account the agreement, to within 20\% of half a period, of the phase of the epoch~3 periodicity extrapolated to epoch~1.  This has a random probability of 0.2.  Thus the probability of the phase agreement in addition to the periodicity agreement between epoch~1 and epoch~3 is $1.58 \times 10^{-6} $, i.e. it is significant at the $4.66 \sigma$ level. This leaves us in no doubt that the periodicity observed in epochs 1 and 3 is a significant physical property of PKS~2131$-$021 and is not simply a random manifestation of red noise.

\begin{figure}[ht!]
\centering
\includegraphics[width=1.0\linewidth]{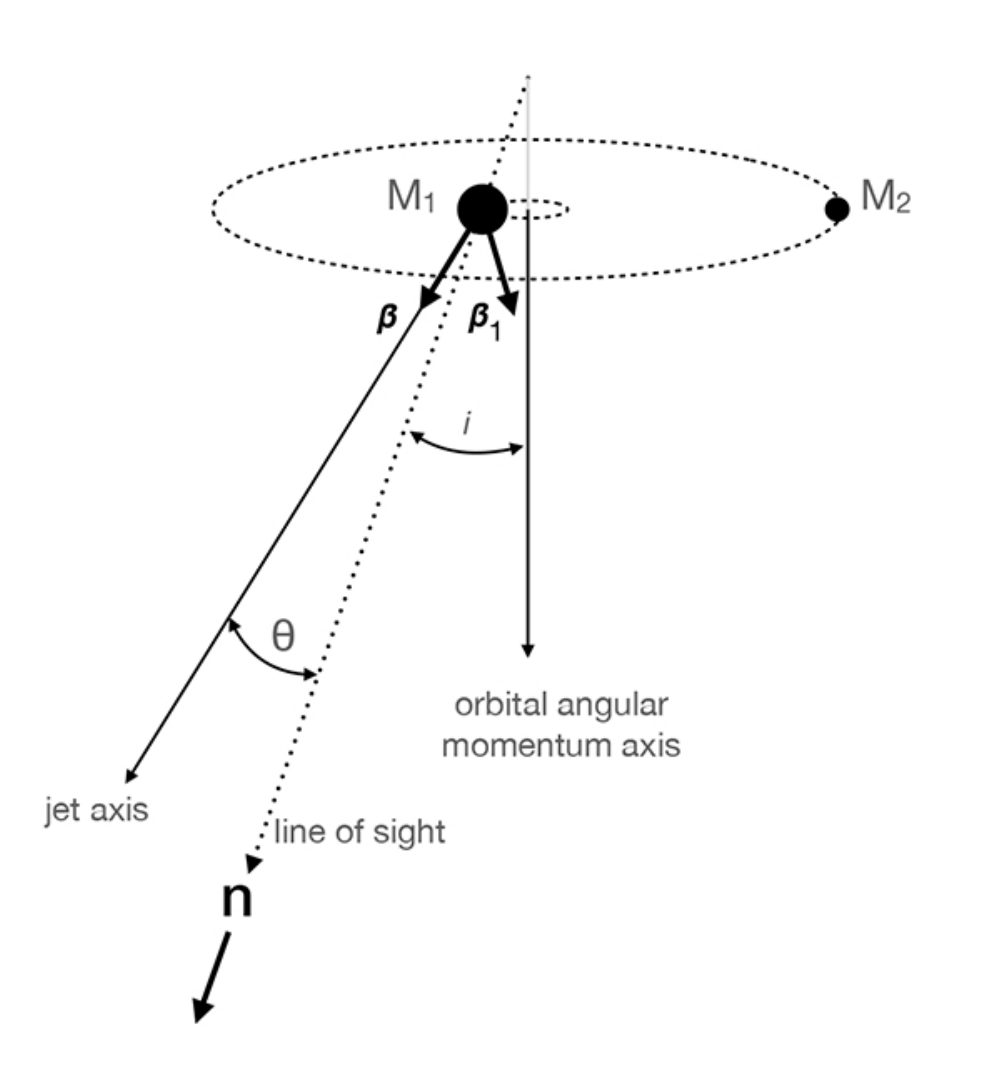}
\caption{Simple model that produces sinusoidal flux density variations in an SMBHB blazar. In the proposed SMBHB, the larger mass, ${\rm M_1}$, and smaller mass, ${\rm M_2}$, orbit the centre of mass with period in the rest frame of the binary,  $P_{\rm }$. We assume that the jet originates in ${\rm M_1}$, although this is not necessary for the model. The angle between the jet axis and the line of sight is $\theta$.    We suppose that the relativistic jet is launched along the  black hole spin axis with constant velocity $c{\bm\beta}$ relative to the black hole.   The orbital velocity  can significantly change the relativistic jet Doppler factor if $\gamma \gg 1$ in the jet, as is the case in PKS~2131$-$021.}
\label{plt:cartoon}
\end{figure}

\section{A Model of PKS~2131-021}\label{sec:model}

We propose the simple model shown in Fig.~\ref{plt:cartoon}, of a black hole binary. We express  the masses of the binary components is units of $10^8M_\odot$, so the primary mass is given by $M_1=10^8M_{1\,8}M_{\odot}$, and similarly for $M_2$.  The secondary, with mass $M_2$, orbits the primary with period in the rest frame of the binary $P=760.6{\rm d}$, and angular momentum that makes an angle $i$ with the line of sight unit vector {\bf n}. We assume that the motion is circular, but elliptical orbits work as well. We assume that the jet is launched along the spin axis of $M_1$ with fixed velocity $c{\bm \beta}$ relative to the black hole. The jet could originate in $M_2$, but this makes no difference to the discussion. The source is a BL Lac object and we know from the above VLBI observations that the line of sight is inclined at a  small angle, $\theta$, to the jet axis.  Let the orbital velocity of $M_1$ be ${\beta}_{\rm  1}=0.036M_{2\,8}/(M_{1\,8}+M_{2\,8})^{2/3}$, where, and henceforth in this section, we set $c=1$.

On our model the orbital motion changes the velocity of the emitting material in the jet relative to the observer, and hence the Doppler factor and beaming. In the case of PKS~2131$-$021 the Doppler factor of the jet is high, since this is a superluminal source with the jet axis closely aligned with the line of sight, as discussed in \S~\ref{sec:vlbi}. In such a case, the orbital motion can have a significant effect on the Doppler factor, as we show below.

Suppose that we have a source at rest emitting isotropically and observed distantly with a flux density $S'$.  The observed flux density will be given by $S= {\cal D} ^{2-\alpha}S'$ \citep{1979Natur.277..182S}, where $\alpha=d\ln S/d\ln\nu$ is the spectral index, and ${\cal D}$ is the Doppler factor:
\begin{equation}
{\cal D} =\frac1{\gamma(1-{\bm\beta}\cdot{\bf n})}
\end{equation}
and $\gamma=(1-\beta^2)^{-1/2}$ is the Lorentz gamma factor.    The orbital motion  causes  both $ \gamma $  and  $\bm\beta$ to change.

Applying the Lorentz transformation from the rest frame of $M_1$ to the rest frame of the binary barycenter, we find

\begin{equation}
\delta{\bm\beta}={\bm\beta}_{\rm 1}-({\bm\beta}\cdot{\bm\beta}_{\rm 1}){\bm\beta}+O(\beta_{\rm 1}^2)
\end{equation}

and we also have
\begin{equation}
\frac{\partial\ln{\cal D}}{\partial{\bm\beta}}=\frac{\bf n}{1-{\bf n}\cdot{\bm\beta}}-\frac{\bm\beta}{1-\beta^2}
\end{equation}
assuming that $\bf n$ is fixed.  
Hence the fractional  change in $S$ is given by
\begin{align}
\delta\ln S&=(2-\alpha) \delta \ln {\cal D}=(2-\alpha)\delta{\bm\beta}\cdot\frac{\partial\ln {\cal D}}{\partial{\bm\beta}}\\
&=\frac{(2-\alpha)({\bf n}-{\bm\beta})\cdot{\bm\beta}_{\rm 1}}{(1-{\bf n}\cdot{\bm\beta})}+O(\beta_{\rm 1}^2).
\end{align}

A continuous jet comprises many such sources, starting and finishing at a supposed fixed rate. Their combined flux density should then satisfy the same relation, to $O(\beta_{\rm 1}^2)$. 

In PKS~2131$-$021, as we have seen in \S~\ref{sec:vlbi}, $\theta\sim\gamma^{-1} \ll 1$. Expanding, we find that $\delta \ln S$ varies sinusoidally
\begin{equation}
\delta \ln S=\frac{2(2-\alpha)\gamma^2\theta \beta_{\rm 1}\cos i}{(1+\gamma^2\theta^2)}\cos(2\pi t/P).
\end{equation}

The observed amplitude, $\delta \ln S \sim0.2$ is compatible with masses $M_{\rm 1\,8}\sim M_{\rm 2\,8}\sim1$ and typical blazar values $\gamma\sim\theta^{-1}\sim10$. 

Intuitively, one would expect the greatest effect on the observed Doppler factor to occur
if the orbital motion is parallel or anti-parallel to the jet, and the least effect to occur if the orbital
motion is perpendicular to the jet. However, as can be seen in equation (7), this is not the case
due to the $({\bm n} - {\bm \beta} )$ term. Since $|n|$ is unity and $\beta$ is very close to unity, ${\bm n} - {\bm \beta} \sim {\bm \theta}$.
Now ${\bm \theta}$ is in the plane of the sky, and hence orthogonal to the line of sight, and it is ${\bm \theta}$ that operates
on ${\bm \beta_1}$, as we see in equation (7). Hence the effect is proportional to ${\rm cos}\, i$, rather then ${\rm sin}\, i$,
contrary to what one would expect intuitively, i.e. a binary whose orbital plane is normal to the line
of sight will give the largest effect,  whereas a binary whose orbital plane lies along the line of sight
will give zero effect to order $O (\beta_1)^2$. 
%  There are two interlinked effects:  (i) The velocity of $M_1$ along the line of sight (in the center of mass frame, which is essentially the observer frame) is maximum when it is approaching us. It causes a relativistic boost to $\beta$, which is sinusoidal and maximum when the line of sight is in the plane of the orbit, i.e. the amplitude of the sinusoid $\propto {\rm sin}\;i$. (ii) The velocity of $M_1$ perpendicular to the line of sight is maximum when it is moving in the plane of the sky. It introduces aberration, essentially changing $\theta$; if $\delta \theta \sim \gamma^{-1}$ , that introduces a sinusoidal change in the received flux density which is out of phase with the first effect and $\propto {\rm cos}\;i$. It so happens that the aberration effect is larger than the velocity along the line of sight effect. For the aberration effect, it is easiest to consider the instantaneous rest frame of $M_1$, i.e. the SMBH producing the jet.  In this instantaneous rest frame it is the earth that is orbiting, and the aberration is simply the usual aberration.  For $M_1=M_2=10^8 M_\odot$ the orbital velocity  $\sim 10^4 {\rm km\, s^{-1}}$, and the total aberration is $\sim 4^\circ$, which is certainly sufficient to produce the observed flux density modulation. 

This result is applicable to the observed variation and is sinusoidal, as measured. However, the quadratic and higher order terms in an expansion will contribute harmonics.  This model is highly simplistic. Real jets are unlikely to be ballistic as they interact with their surroundings. They can accelerate and decelerate. Also the emission is quite likely to originate from radii which are not much smaller than $cP$, introducing retarded time effects which will also lead to higher harmonics.  

However this simple model demonstrates that with plausible values of the black hole masses and jet speed, it is possible to account for the sinusoidal flux density variations observed in PKS~2131$-$021. Numerical simulations of jets can be used to explore the observed behavior of orbiting jets more carefully.

As discussed in \S~\ref{sec:large}, precession has been suggested to explain  the unusual large scale radio structure of PKS~2131$-$021. \cite{1980Natur.287..307B} show that in an SMBHB the more massive component will undergo geodetic precession with period
\begin{equation}
P_{\rm prec} \sim  400 \;r_{16} \left(\frac{M_1}{M_2}\right) \frac{P}{M_{1\,8}}
\end{equation}
in which the orbit is assumed to be circular, and where $r_{16}$ is the separation of the binary in units of $10^{16}$ cm. In PKS~2131$-$021 $r_{16}\sim 1$ for $M_{1\,8}=1$ (see equation \ref{eqn:separation}).

It is therefore entirely possible, and even likely, that the unusual large scale radio structure in PKS~2131$-$021  is due to geodetic precession, while the radio and infrared light curve periodicity is due to orbital motion. 

From Figs. \ref{plt:composite}(a), (b), and (c), we measure the opening angle of the jet to be $73^\circ \pm 18^\circ$. As discussed in \S~\ref{sec:vlbi}, the viewing angle in PKS~2131$-$021 is $\theta=3.8^\circ$ \citep{2021arXiv210904977H}. Thus the precession cone opening angle would be $4.8^\circ \pm 1.2^\circ$, which would seem entirely reasonable.

\begin{figure*}[!t]
\centering
\includegraphics[width=\textwidth]{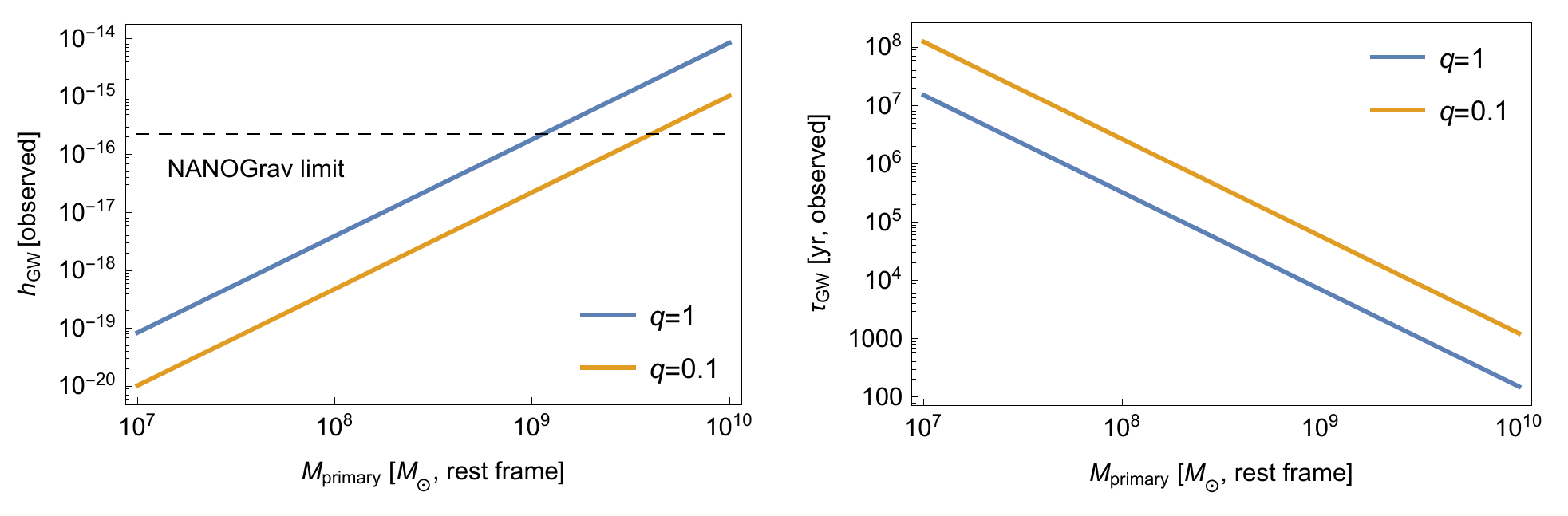}
\caption{Mass limits from GW observations. Both the period and the location of PKS~2131$-$021 are in the windows of maximum sensitivity for the NANOGrav pulsar timing array \citep{2019ApJ...880..116A}. Here we show the observed GW strain and time to coalescence for primary-SMBH rest masses ranging from $10^7 M_\odot$ to $10^{10} M_\odot$, and for binary mass ratios of $q=1.0$ and $q=0.1$.}
\label{plt:gravstrain}
\end{figure*}

A more detailed model for the nature of the activity observed in PKS~2131$-$021 goes beyond the scope of this paper, but should be able to reproduce the following features of the light curve:

\noindent
(a) periodic variations with period observed on Earth $P_\oplus=1737.9\pm2.6$~days are episodic and dominate the light curve at times, and are at other times invisible.  As mentioned in the introduction,  it is not difficult to invent models in which the sinusoidal signal switches on and off.  However the appearance and disappearance of the sinusoidal signal is an added complexity that would have to be accommodated in any complete model of the PKS 2131-021 system. \\ 
(b) When the periodic behavior is observed, it manifests itself with a remarkable stability of period and phase. \\
(c) The flux density is lower  when the periodic behavior is absent.  \\
(d) The amplitude of the periodic behavior, during different epochs of its manifestation, can be very different.

All these features can be naturally explained assuming that we are observing the superposition of the output from two distinct emission processes: 

\noindent
(I) periodic, originating in a part of the jet directly affected by the SMBHB, \\
(II) non-periodic, originating over a wide range of locations along the jet. 
Process (II) is the usual mode of activity seen in VLBI-resolved emission in typical blazar jets: strongly variable, largely stochastic,  as shown by the sinusoid-subtracted data in Fig.~\ref{plt:light curves 3}.   Features (c) and (d) of the light curve, pointed out above, strongly suggest that process (I) is also strongly variable in radiative output. If this were not the case, it would be hard to explain how it is possible for the periodicity to disappear when the overall flux density decreases. The qualitative features of the summed light curve during different epochs clearly depend on the relative level of activity between the two processes:

\noindent - If both processes (I) and (II) are in a high state, we see a noisy periodic signal, as in epoch~1. \\ 
- If process (I) is in a low state, we see only aperiodic variations produced by process (II). The flux density level depends on the level of activity of (II), as seen in epoch~2\\
-  If emission due to process (II) is low, or stable, while process (I) is in its high state, then we observe a strongly periodic signal with little noise, as seen in epoch~3.

\section{Implications for gravitational-wave emission}
\label{sec:nanograv}

The characteristic gravitational-wave (GW) strain produced by a circular SMBHB, averaged over binary inclination, is \citep{2019ApJ...880..116A,2021ApJ...914..121A}
\begin{equation}
h_{\mathrm{GW}} = \frac{2(\mathrm{G}{\cal{M}_\oplus})^{5/3}(\pi f_{\mathrm{GW\oplus}})^{2/3}}{\mathrm{c}^4D_L},
\label{eqn:strain}
\end{equation}
where $\cal{M}_\oplus$ is the ``chirp mass'' in the observer frame, $D_L$ is the luminosity distance, $f_{\mathrm{GW\oplus}}$ is the GW frequency as observed on Earth, and~$\mathrm{G}$ and~$\mathrm{c}$ are the standard physical constants.
The observed chirp mass~$\cal{M}_\oplus$ is related to the rest-frame chirp mass $\mathcal{M}_{\rm }$ and individual masses $M_1$, $M_2$ by \citep{maggiore2008}
\begin{equation}
\mathcal{M}_{\rm } = \frac{(M_1 M_2)^{3/5}}{(M_1 + M_2)^{1/5}} = \frac{\mathcal{M}_\oplus}{1+z}.
\label{eqn:strain1}
\end{equation}
The observed GW frequency $f_{\mathrm{GW\oplus}}$ and the observed orbital period of the binary~$P_\oplus$ are related as $f_{\mathrm{GW\oplus}} = 2/P_\oplus$.
Adopting values appropriate for PKS~2131$-$021, namely $D_L = 9.08$~Gpc and $f_{\rm GW\oplus} = 1.3 \times 10^{-8}$~Hz, and assuming that the binary is ``face on,'' we find 
\begin{equation}
h_{\mathrm{GW}}=1.5 \times 10^{-16} (\mathcal{M}_{\rm } /10^9\,M_\odot)^{5/3},
\label{eqn:strain2}
\end{equation}
while the rest-frame separation of the binary is
\begin{equation}
r = \left(\frac{G m_{\rm }}{4\pi^2}\right)^{1/3}
\left(\frac{P_\oplus}{1+z}\right)^{2/3} \simeq
9.6 \left(\frac{m_{\rm }}{10^9 M_\odot}\right)^{1/3} \mathrm{light \; days}
\label{eqn:separation}
\end{equation}
where $m_{\rm }$ is the total rest-frame mass. Thus, for $m\sim 3 \times 10^6M_\odot - 3 \times 10^9M_\odot$, we have $r \sim 0.001 - 0.01$ pc.

The PKS~2131$-$021 period is well matched to the sensitivity window of current GW searches with pulsar timing arrays, and its sky location is almost optimal for the NANOGrav array, which reported a 95\% Bayesian upper limit on characteristic strain $h_{\rm GW} \lesssim 2.25 \times 10^{-16}$ \citep{2019ApJ...880..116A}. This translates to a limit $\mathcal{M}_{\rm } \lesssim 4 \times 10^9 M_\odot$ on rest-frame chirp mass, or $\lesssim 9 \times 10^9 M_\odot$ on rest-frame \emph{total} mass for equal-mass components (see also Fig.~\ \ref{plt:gravstrain}).

If PKS~2131$-$021 is indeed an SMBHB with such a high chirp mass, it will be within detection range for the joint dataset of the International Pulsar Timing Array, enhanced with pulsars found in planned surveys in the first few years of the Square Kilometer Array \citep{2021ApJ...915...97X}. 
However, it is interesting to consider the likelihood that we would happen to observe an SMBHB in this particular stage of its evolution. GW strain and (observed) time to coalescence $t_\oplus$ are related by
\begin{equation}
h_{\mathrm{GW}}=\frac{5 \, c \, P_\oplus^2}{512\,\pi^2 \, t_\oplus \, D_L},
\label{eqn:strain3}
\end{equation}
independent of chirp mass, as can be derived, for instance, using the equations in \cite{maggiore2008}. Evaluating Eq.\ \eqref{eqn:strain3} for PKS~2131$-$021, we obtain
\begin{equation}
h_{\mathrm{GW}} \simeq 8 \times 10^{-19} \; (t_\oplus / \mathrm{Myr})^{-1};
\label{eqn:strain4}
\end{equation}
equating expressions \eqref{eqn:strain2} and \eqref{eqn:strain4} yields $t_\oplus \sim 5000$ yr for $\mathcal{M}_{\rm } = 10^9 M_\odot$, which implies that the higher chirp masses allowed by GW limits would require implausible observational serendipity. However, this is mitigated by the fact that PKS~2131$-$021 is by far the best SMBHB candidate we have found in the sample of  $\sim 1830$ blazars that we have been monitoring for 13 years at the OVRO.  In this regard we note that \cite{2018MNRAS.481L..74H} estimate, from mock population studies based on the luminosity functions for BL Lacertae objects and flat-spectrum radio quasars with redshifts $z \le 2$, that a fraction $\le 10^{-3}$ of blazars host a binary with an orbital period $P<5$ yr, which is not inconsistent with our statistic here of one strong SMBHB candidate out of a sample of $\sim 1800$ blazars. It should be clear that adding more years of OVRO data is essential for identifying other examples which may have longer periods, or which may have been in a phase of not showing sinusoidal variations during the 13-year OVRO time window.

\section{Discussion}\label{sec:conclusions}

We have shown that PKS~2131$-$021 exhibits unique periodicity behavior over a 45.1-year observing span: two epochs of periodic emission, separated by 20 years, agree in both period and phase. This periodic feature in the PKS~2131$-$021 light curve is significant at the $4.6\sigma$ level. This is strongly suggestive of an SMBHB.
We have also shown that,   due to red noise,  small $\sim 10\%$ variations in the periodicity are easily explained and are, in fact, to be expected in the light curves of blazars.
The form of the periodic variability in epochs 1 and 3 is unexpectedly sinusoidal in character and this must be telling us something important about the source. We have demonstrated in a simple model that this can be accounted for by the Doppler effect of  orbital motion in an SMBHB blazar due to the strong effect of the orbital motion on the Doppler factor of the relativistic jet.

\vskip 2pt
\noindent
{\it Occam's Razor:} While we have not yet proven definitively that PKS 2131-021 is an SMBHB, we believe that this is by far the most likely scenario. We know that PKS 2131-021 has a highly relativistic jet oriented close to the line of sight, which exhibits random flux density variations on timescales of months to years. If it is an SMBHB then we also know that the SMBH generating the jet is in orbital motion. We have shown that this will produce a sinusoidal modulation of the observed flux density of approximately the magnitude observed.  It is also highly likely that features appear and disappear in the light curves of blazars as a result of fuelling of the central engine.  Therefore, the economy of requiring no additional assumptions, other than that of being an SMBHB, to explain all the observations in PKS 2131-021 satisfies Occam’s razor.

If PKS~2131$-$021 is indeed an SMBHB, it is of interest for the constraints it could provide on models of SMBHB merger progenitors \citep{2021NatAs...5..569S}. If sufficiently massive, which appears unlikely, this SMBHB would be a future candidate for gravitational-wave detection with pulsar timing arrays, which can already provide an upper limit on its chirp mass.
Its confirmation as a binary would therefore be consequential, and it can be expected within the next 5--10 years from continuing light-curve observations with the 40 m Telescope of the OVRO, provided the periodic variations continue as they have over the past 16 years. 

 By far the most important conclusions of this paper are (i) regardless of whether or not PKS~2131$-$021 is an SMBHB, sinusoidal flux density fluctuations, due to orbital motion, and the inevitable~apparent variations in period caused by red noise, are observational properties of supermassive black hole binaries with relativistic jets that clearly should be anticipated; (ii) we should expect to see gaps in the sinusoidal variations, possibly lasting decades; and (iii) in this field the observed phenomenology must lead, rather than the theory,  because the detailed models that are required cannot be predicted by theory.  Thus this object provides an extremely useful blueprint for the analysis of SMBHB candidate light curves, including accounting for the effects of random variability that is not associated with the periodicity of interest. We hope that it provides a useful demonstration of the complementary advantages of the GLS, WWZ, and least-squares sine wave fitting analyses for the investigation of light curves in the pursuit of SMBHBs. We trust that these legacy results from Haystack, UMRAO, and OVRO have convinced the reader of the importance of long term monitoring in astronomy, and its potential for making discoveries.

\section*{Acknowledgements}

We thank the anonymous referee for carrying out a very detailed and thorough review of the original manuscript and for many insightful suggestions and questions, which have greatly improved this paper.  We thank the California Institute of Technology and the Max Planck Institute for Radio Astronomy for supporting the  OVRO 40\,m program under extremely difficult circumstances over the last 5 years in the absence of agency funding. Without this private support this paper would not have been written.  We also thank all the volunteers who have enabled this work to be carried out.   Prior to 2016, the OVRO program was supported by NASA grants NNG06GG1G, NNX08AW31G, NNX11A043G, and NNX13AQ89G from 2006 to 2016 and NSF grants AST-0808050, and AST-1109911 from 2008 to 2014. The UMRAO program received support from NSF grants AST-8021250, AST-8301234, AST-8501093, AST-8815678, AST-9120224, AST-9421979, AST-9617032, AST-9900723, AST-0307629, AST-0607523, and earlier NSF awards, and from NASA grants NNX09AU16G, NNX10AP16G, NNX11AO13G, and NNX13AP18G. Additional funding for the operation of UMRAO was provided by the University of Michigan. The NANOGrav project receives support from National Science Foundation (NSF) Physics Frontier Center award number 1430284. T. H. was supported by the Academy of Finland projects 317383, 320085, and 322535. S.K. acknowledges support from the European Research Council (ERC) under the European Unions Horizon 2020 research and innovation programme under grant agreement No.~771282. W.M. acknowledges support from ANID projects Basal AFB-170002 and FONDECYT 11190853. R.R. acknowledges support from ANID Basal AFB-170002, and ANID-FONDECYT grant 1181620.  C.O. acknowledges support from the Natural Sciences and Engineering Research Council (NSERC) of Canada. M.V.\ and T.J.W.L.\ performed part of this work at the Jet Propulsion Laboratory, California Institute of Technology, under a contract with the National Aeronautics and Space Administration (80NM0018D0004). V.P. acknowledges support from the Foundation of Research and Technology - Hellas Synergy Grants Program through project MagMASim, jointly implemented by the Institute of Astrophysics and the Institute of Applied and Computational Mathematics and by the Hellenic Foundation for Research and Innovation (H.F.R.I.) under the “First Call for H.F.R.I. Research Projects to support Faculty members and Researchers and the procurement of high-cost research equipment grant” (Project 1552 CIRCE).  S.O. gratefully acknowledges the support of
the Caltech Summer Undergraduate Research Fellowship program.

\appendix

\section{Estimating the significance of periodicities in blazar light curves}
\label{app:significance}

\begin{figure}[!t]
\centering
\includegraphics[width=1.0\linewidth]{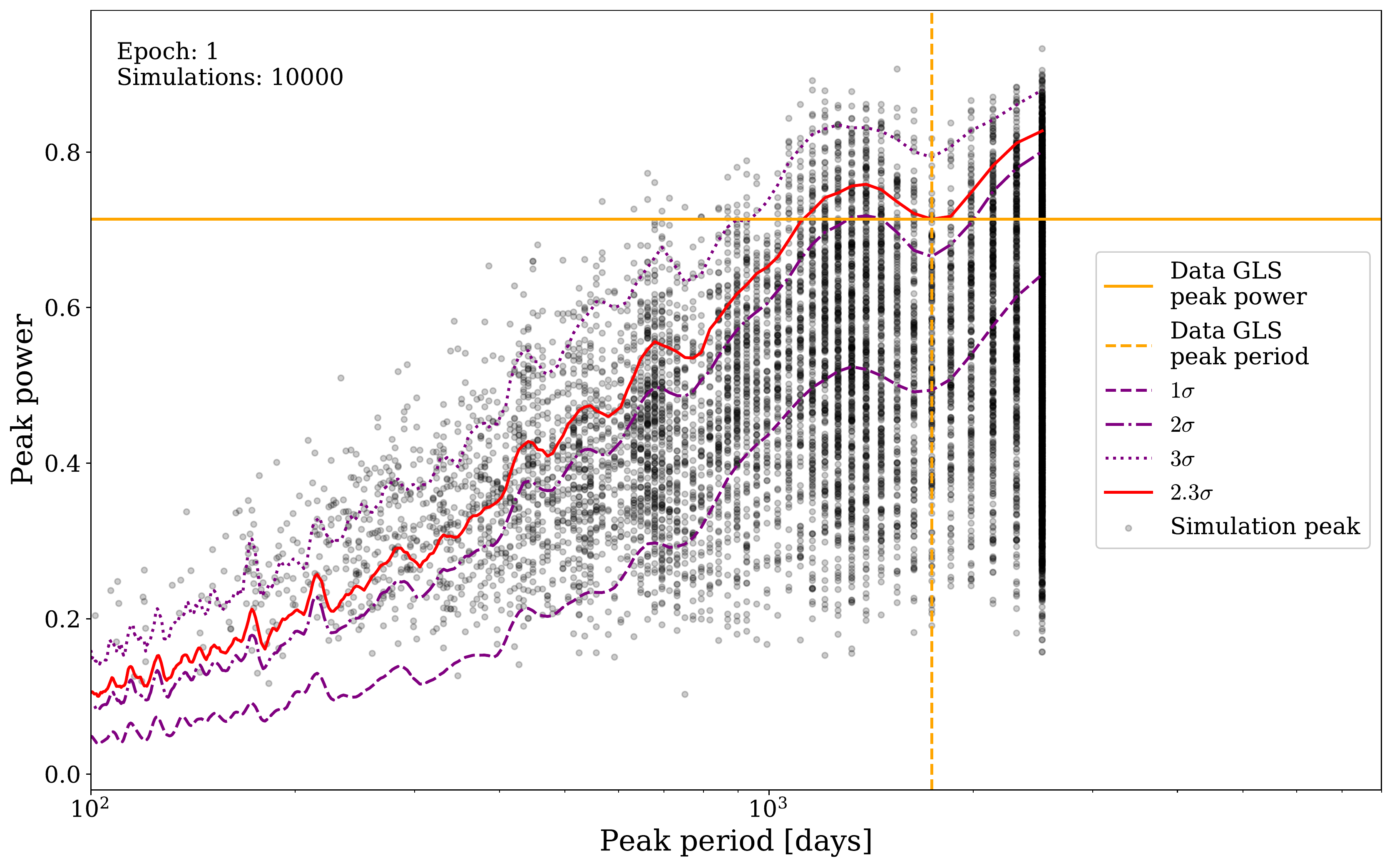}
\caption{ Calculation of  the global p-value using epoch~1 as an example.  The quantization of the periods is due to the quantization in frequency, $\Delta f = 1/(\zeta T)$, with $\zeta= 10$, which we impose in running the GLS.  So the sampled periods are  separated  by $1/(10 \Delta f)$.  Step~1: Orange vertical and horizontal lines mark the period and power, respectively, of the strongest peak identified in the GLS of the epoch~1 data.   The curves plotted here are merely a visualization tool, they do not represent levels of significance {\it unless} there are {\it a priori} reasons for selecting particular periods. Hence in the next step (Step 2), 
the detection of the strongest peak in the GLS of the data at the period $P_{\rm peak}$ is a  $2.3\sigma$~event only in the case that there is an {\it a priori} reason for selecting this particular period. The significance is  obtained by summing the simulations with $p-$values less than that observed at this particular period. The red line connects the $2.3\sigma$~significance level at each period.
The purple curves indicate the $1\sigma, \; 2\sigma, $ and $3\sigma$ significance levels at each specific period, as also shown in the top panel of Fig.~\ref{plt:LSpowers}, which are misleading unless there is an {\it a priori} reason for selecting a specific period.
Step~3: For each simulation the strongest peak is identified in the GLS. The peak periods and powers are shown as grey dots.
Step~4: To estimate the global  p-value,  simulations at all periods are counted in which the power of the strongest peak lies above the $2.3\sigma$~significance level (red line). }
\label{plt:LSpeaks}
\end{figure}

The flaring behaviour of blazars can lead to spurious periodicity ``detections''. It is therefore important to make reliable estimates of the significance of any likely detections. Here we suggest an approach for avoiding the known pitfalls of red noise and single epoch~significance tests.

% simulations:
For each epoch~we simulate light curves that match the power spectral density (PSD), the probability density function (PDF), sampling and observational noise of the specific epoch.
Assuming a pure power-law $\mathrm{PSD} \sim \nu^{-\beta}$, where $\nu$ is the frequency, we use our own implementation of the method introduced by \citet{2002MNRAS.332..231U} to estimate the following power law slopes for epochs 1--3:
$\beta_{\rm epoch~1}=1.71\pm0.19$, $\beta_{\rm epoch~2}=1.75\pm0.26$, $\beta_{\rm epoch~3}=1.82\pm0.14$, showing no significant difference between the three epochs.
An Anderson-Darling test shows that the PDF of each epoch~is not Gaussian distributed at significance level $<10^{-3}$.
Therefore, we use the algorithm of \citet{2013MNRAS.433..907E} to create artificial light curves that match the estimated PSDs and PDFs. The initial simulations are sampled such that low- and high-frequency power is included beyond the  data ranges. The simulations are then re-sampled to the time stamps of the data and Gaussian noise is added based on the estimated flux density uncertainties.

The \emph{single-period p-value} is the probability of finding a periodic signal at the observed frequency and with the same or greater power under  the null  hypothesis that the signal is the result of a red-noise process with the same PSD, PDF, sampling, and observational noise as the real data. 
We use the following procedure { -- illustrated for epoch~1 in Fig.~\ref{plt:LSpeaks} --} to estimate the \emph{global p-value}, i.e. the probability of finding a periodic signal under the above null hypothesis as significant as the one observed at any frequency:

\begin{enumerate}
\item We calculate the GLS periodogram of the data, select the strongest peak and measure its peak period, $P_{\rm peak}$, and power, $\mathcal{P}_{\rm peak}$.
\item At period $P_{\rm peak}$ we count all simulations with power $\mathcal{P}_{\rm sim} \ge \mathcal{P}_{\rm peak}$ to estimate the single-period p-value, $p_{\rm peak}$.
\item For each simulation,
\begin{enumerate}
    \item we calculate the GLS periodogram at the same discrete frequencies as for the data, we select the strongest peak and measure its peak period, $P_{\rm sim}$, and power, $\mathcal{P}_{\rm sim}$.
    \item At $P_{\rm sim}$ we calculate the single-period p-value, $p_{\rm sim}$.
\end{enumerate}
\item We count all simulations for which $p_{\rm sim} \le p_{\rm peak}$ to estimate the global p-value.
\end{enumerate}

As explained in \S~\ref{sec:caution}, the single period p-value is commonly used to express the significance of a detected periodicity. However, it is an incorrect and misleading estimate of the significance, unless this period has been selected {\it a priori}, since it does not take into account the fact that spurious periodicities from a red-noise process could arise at any period, and not just the one detected. 

We report the global p-values estimated for epochs 1--3 in Table~\ref{tab:lsvalues} as Tests 1.1, 1.2, and 1.3. Test 1.3, is used in the additional significance estimates, starting from this highly significant detection, described in Sect.~\ref{sec:sign1}.

\section{The WWZ Transform}
\label{app:wwz}

Following \cite{1996AJ....112.1709F}, the WWZ transform 
is defined here as:

\begin{equation}
    WWZ=\frac{(N_{eff}-3)V_y}{2(V_x-V_y)}
\end{equation}

where $N_{eff}$ is the effective number of data points, and $V_x$ and $V_y$ are the weighted variances of the data and the model, respectively, as defined by  \cite{1996AJ....112.1709F}.
For PKS 2131-021, we found that the $c$-value of 0.125 proposed by Foster was a good choice, since other values  degraded either the resolution of the time or the period. The density of data points, and hence $N_{eff}$, varies strongly between the three epochs (see Fig. \ref{plt:light curves}).  Also, as can be seen in equation (B1), when the data and model variances approach each other, which happens when the data approach a sinusoidal waveform, the denominator approaches zero and hence the WWZ power increases. This is why the WWZ magnitude in epoch $3$ seen in the upper panel of Fig. 8 reaches a high value ($1790$). This is $\sim 20-30 \times$  higher than the peak WWZ power in epoch 1. The signal in epoch 1 is noisier, with higher variance, and therefore the WWZ power is considerably smaller than in epoch 3. However, although the difference in power for the two epochs is large, the peak in epoch 1 is highly relevant since its period agrees with that of epoch 3 to within $\sim 2\%$, and in addition, as we have learned from the least squares sine wave fitting, the phase of the epoch 1 periodicity agrees, to within 10\% of the periodicity period, with the extrapolated phase from epoch 3.

\newpage

\bibliography{sample63}{}

\begin{thebibliography}{}
\expandafter\ifx\csname natexlab\endcsname\relax\def\natexlab#1{#1}\fi
\providecommand{\url}[1]{\href{#1}{#1}}
\providecommand{\dodoi}[1]{doi:~\href{http://doi.org/#1}{\nolinkurl{#1}}}
\providecommand{\doeprint}[1]{\href{http://ascl.net/#1}{\nolinkurl{http://ascl.net/#1}}}
\providecommand{\doarXiv}[1]{\href{https://arxiv.org/abs/#1}{\nolinkurl{https://arxiv.org/abs/#1}}}

\bibitem[{{Abraham}(2018)}]{2018NatAs...2..443A}
{Abraham}, Z. 2018, Nature Astronomy, 2, 443, \dodoi{10.1038/s41550-018-0484-2}

\bibitem[{{Aggarwal} {et~al.}(2019){Aggarwal}, {Arzoumanian}, {Baker},
  {Brazier}, {Brinson}, {Brook}, {Burke-Spolaor}, {Chatterjee}, {Cordes},
  {Cornish}, {Crawford}, {Crowter}, {Cromartie}, {DeCesar}, {Demorest},
  {Dolch}, {Ellis}, {Ferdman}, {Ferrara}, {Fonseca}, {Garver-Daniels},
  {Gentile}, {Hazboun}, {Holgado}, {Huerta}, {Islo}, {Jennings}, {Jones},
  {Jones}, {Kaiser}, {Kaplan}, {Kelley}, {Key}, {Lam}, {Lazio}, {Levin},
  {Lorimer}, {Luo}, {Lynch}, {Madison}, {McLaughlin}, {McWilliams},
  {Mingarelli}, {Ng}, {Nice}, {Pennucci}, {Pol}, {Ransom}, {Ray}, {Siemens},
  {Simon}, {Spiewak}, {Stairs}, {Stinebring}, {Stovall}, {Swiggum}, {Taylor},
  {Turner}, {Vallisneri}, {van Haasteren}, {Vigeland}, {Witt}, {Zhu}, \&
  {NANOGrav Collaboration}}]{2019ApJ...880..116A}
{Aggarwal}, K., {Arzoumanian}, Z., {Baker}, P.~T., {et~al.} 2019, \apj, 880,
  116, \dodoi{10.3847/1538-4357/ab2236}

\bibitem[{{Aller} {et~al.}(1985){Aller}, {Aller}, {Latimer}, \&
  {Hodge}}]{1985ApJS...59..513A}
{Aller}, H.~D., {Aller}, M.~F., {Latimer}, G.~E., \& {Hodge}, P.~E. 1985,
  \apjs, 59, 513, \dodoi{10.1086/191083}

\bibitem[{{Aller} {et~al.}(1999){Aller}, {Aller}, {Hughes}, \&
  {Latimer}}]{1999ApJ...512..601A}
{Aller}, M.~F., {Aller}, H.~D., {Hughes}, P.~A., \& {Latimer}, G.~E. 1999,
  \apj, 512, 601, \dodoi{10.1086/306799}

\bibitem[{{Arzoumanian} {et~al.}(2021){Arzoumanian}, {Baker}, {Brazier},
  {Brook}, {Burke-Spolaor}, {Becsy}, {Charisi}, {Chatterjee}, {Cordes},
  {Cornish}, {Crawford}, {Cromartie}, {Decesar}, {Demorest}, {Dolch},
  {Elliott}, {Ellis}, {Ferrara}, {Fonseca}, {Garver-Daniels}, {Gentile},
  {Good}, {Hazboun}, {Islo}, {Jennings}, {Jones}, {Kaiser}, {Kaplan}, {Kelley},
  {Key}, {Lam}, {Lazio}, {Luo}, {Lynch}, {Ma}, {Madison}, {McLaughlin},
  {Mingarelli}, {Ng}, {Nice}, {Pennucci}, {Pol}, {Ransom}, {Ray},
  {Shapiro-Albert}, {Siemens}, {Simon}, {Spiewak}, {Stairs}, {Stinebring},
  {Stovall}, {Swiggum}, {Taylor}, {Vallisneri}, {Vigeland}, {Witt}, \&
  {Nanograv Collaboration}}]{2021ApJ...914..121A}
{Arzoumanian}, Z., {Baker}, P.~T., {Brazier}, A., {et~al.} 2021, \apj, 914,
  121, \dodoi{10.3847/1538-4357/abfcd3}

\bibitem[{{Baars} {et~al.}(1977){Baars}, {Genzel}, {Pauliny-Toth}, \&
  {Witzel}}]{1977A&A....61...99B}
{Baars}, J.~W.~M., {Genzel}, R., {Pauliny-Toth}, I.~I.~K., \& {Witzel}, A.
  1977, \aap, 500, 135

\bibitem[{{Balonek}(1982)}]{1982PhDT.........6B}
{Balonek}, T.~J. 1982, PhD thesis, Massachusetts Univ., Amherst.

\bibitem[{{Bansal} {et~al.}(2017){Bansal}, {Taylor}, {Peck}, {Zavala}, \&
  {Romani}}]{2017ApJ...843...14B}
{Bansal}, K., {Taylor}, G.~B., {Peck}, A.~B., {Zavala}, R.~T., \& {Romani},
  R.~W. 2017, \apj, 843, 14, \dodoi{10.3847/1538-4357/aa74e1}

\bibitem[{{Begelman} {et~al.}(1980){Begelman}, {Blandford}, \&
  {Rees}}]{1980Natur.287..307B}
{Begelman}, M.~C., {Blandford}, R.~D., \& {Rees}, M.~J. 1980, \nat, 287, 307,
  \dodoi{10.1038/287307a0}

\bibitem[{{Britzen} {et~al.}(2018){Britzen}, {Fendt}, {Witzel}, {Qian},
  {Pashchenko}, {Kurtanidze}, {Zajacek}, {Martinez}, {Karas}, {Aller}, {Aller},
  {Eckart}, {Nilsson}, {Ar{\'e}valo}, {Cuadra}, {Subroweit}, \&
  {Witzel}}]{2018MNRAS.478.3199B}
{Britzen}, S., {Fendt}, C., {Witzel}, G., {et~al.} 2018, \mnras, 478, 3199,
  \dodoi{10.1093/mnras/sty1026}

\bibitem[{{Burke-Spolaor} {et~al.}(2019){Burke-Spolaor}, {Taylor}, {Charisi},
  {Dolch}, {Hazboun}, {Holgado}, {Kelley}, {Lazio}, {Madison}, {McMann},
  {Mingarelli}, {Rasskazov}, {Siemens}, {Simon}, \&
  {Smith}}]{2019A&ARv..27....5B}
{Burke-Spolaor}, S., {Taylor}, S.~R., {Charisi}, M., {et~al.} 2019, \aapr, 27,
  5, \dodoi{10.1007/s00159-019-0115-7}

\bibitem[{{Caproni} {et~al.}(2017){Caproni}, {Abraham}, {Motter}, \&
  {Monteiro}}]{2017ApJ...851L..39C}
{Caproni}, A., {Abraham}, Z., {Motter}, J.~C., \& {Monteiro}, H. 2017, \apjl,
  851, L39, \dodoi{10.3847/2041-8213/aa9fea}

\bibitem[{{Caproni} {et~al.}(2004){Caproni}, {Mosquera Cuesta}, \&
  {Abraham}}]{2004ApJ...616L..99C}
{Caproni}, A., {Mosquera Cuesta}, H.~J., \& {Abraham}, Z. 2004, \apjl, 616,
  L99, \dodoi{10.1086/426863}

\bibitem[{{Cassaro} {et~al.}(2002){Cassaro}, {Stanghellini}, {Dallacasa},
  {Bondi}, \& {Zappal{\`a}}}]{2002A&A...381..378C}
{Cassaro}, P., {Stanghellini}, C., {Dallacasa}, D., {Bondi}, M., \&
  {Zappal{\`a}}, R.~A. 2002, \aap, 381, 378, \dodoi{10.1051/0004-6361:20011460}

\bibitem[{{Covino} {et~al.}(2019){Covino}, {Sandrinelli}, \&
  {Treves}}]{2019MNRAS.482.1270C}
{Covino}, S., {Sandrinelli}, A., \& {Treves}, A. 2019, \mnras, 482, 1270,
  \dodoi{10.1093/mnras/sty2720}

\bibitem[{{Dent} {et~al.}(1974{\natexlab{a}}){Dent}, {Aller}, \&
  {Olsen}}]{1974ApJ...188L..11D}
{Dent}, W.~A., {Aller}, H.~D., \& {Olsen}, E.~T. 1974{\natexlab{a}}, \apjl,
  188, L11, \dodoi{10.1086/181418}

\bibitem[{{Dent} {et~al.}(1974{\natexlab{b}}){Dent}, {Kapitzky}, \&
  {Kojoian}}]{1974AJ.....79.1232D}
{Dent}, W.~A., {Kapitzky}, J.~E., \& {Kojoian}, G. 1974{\natexlab{b}}, \aj, 79,
  1232, \dodoi{10.1086/111665}

\bibitem[{{Dey} {et~al.}(2021){Dey}, {Valtonen}, {Gopakumar}, {Lico},
  {G{\'o}mez}, {Susobhanan}, {Komossa}, \& {Pihajoki}}]{2021MNRAS.503.4400D}
{Dey}, L., {Valtonen}, M.~J., {Gopakumar}, A., {et~al.} 2021, \mnras, 503,
  4400, \dodoi{10.1093/mnras/stab730}

\bibitem[{{Drake} {et~al.}(2009){Drake}, {Djorgovski}, {Mahabal}, {Beshore},
  {Larson}, {Graham}, {Williams}, {Christensen}, {Catelan}, {Boattini},
  {Gibbs}, {Hill}, \& {Kowalski}}]{drake2009}
{Drake}, A.~J., {Djorgovski}, S.~G., {Mahabal}, A., {et~al.} 2009, \apj, 696,
  870, \dodoi{10.1088/0004-637X/696/1/870}

\bibitem[{{Drinkwater} {et~al.}(1997){Drinkwater}, {Webster}, {Francis},
  {Condon}, {Ellison}, {Jauncey}, {Lovell}, {Peterson}, \&
  {Savage}}]{1997MNRAS.284...85D}
{Drinkwater}, M.~J., {Webster}, R.~L., {Francis}, P.~J., {et~al.} 1997, \mnras,
  284, 85, \dodoi{10.1093/mnras/284.1.85}

\bibitem[{{Emmanoulopoulos} {et~al.}(2013){Emmanoulopoulos}, {McHardy}, \&
  {Papadakis}}]{2013MNRAS.433..907E}
{Emmanoulopoulos}, D., {McHardy}, I.~M., \& {Papadakis}, I.~E. 2013, \mnras,
  433, 907, \dodoi{10.1093/mnras/stt764}

\bibitem[{{Emmanoulopoulos} {et~al.}(2010){Emmanoulopoulos}, {McHardy}, \&
  {Uttley}}]{2010MNRAS.404..931E}
{Emmanoulopoulos}, D., {McHardy}, I.~M., \& {Uttley}, P. 2010, \mnras, 404,
  931, \dodoi{10.1111/j.1365-2966.2010.16328.x}

\bibitem[{{Foreman-Mackey} {et~al.}(2013){Foreman-Mackey}, {Hogg}, {Lang}, \&
  {Goodman}}]{Foreman2013}
{Foreman-Mackey}, D., {Hogg}, D.~W., {Lang}, D., \& {Goodman}, J. 2013, \pasp,
  125, 306, \dodoi{10.1086/670067}

\bibitem[{{Foster}(1996)}]{1996AJ....112.1709F}
{Foster}, G. 1996, \aj, 112, 1709, \dodoi{10.1086/118137}

\bibitem[{{Graham} {et~al.}(2015{\natexlab{a}}){Graham}, {Djorgovski}, {Stern},
  {Drake}, {Mahabal}, {Donalek}, {Glikman}, {Larson}, \&
  {Christensen}}]{2015MNRAS.453.1562G}
{Graham}, M.~J., {Djorgovski}, S.~G., {Stern}, D., {et~al.} 2015{\natexlab{a}},
  \mnras, 453, 1562, \dodoi{10.1093/mnras/stv1726}

\bibitem[{{Graham} {et~al.}(2015{\natexlab{b}}){Graham}, {Djorgovski}, {Stern},
  {Glikman}, {Drake}, {Mahabal}, {Donalek}, {Larson}, \&
  {Christensen}}]{2015Natur.518...74G}
---. 2015{\natexlab{b}}, \nat, 518, 74, \dodoi{10.1038/nature14143}

\bibitem[{{Graham} {et~al.}(2019){Graham}, {Kulkarni}, {Bellm}, {Adams},
  {Barbarino}, {Blagorodnova}, {Bodewits}, {Bolin}, {Brady}, {Cenko}, {Chang},
  {Coughlin}, {De}, {Eadie}, {Farnham}, {Feindt}, {Franckowiak}, {Fremling},
  {Gezari}, {Ghosh}, {Goldstein}, {Golkhou}, {Goobar}, {Ho}, {Huppenkothen},
  {Ivezi{\'c}}, {Jones}, {Juric}, {Kaplan}, {Kasliwal}, {Kelley}, {Kupfer},
  {Lee}, {Lin}, {Lunnan}, {Mahabal}, {Miller}, {Ngeow}, {Nugent}, {Ofek},
  {Prince}, {Rauch}, {van Roestel}, {Schulze}, {Singer}, {Sollerman}, {Taddia},
  {Yan}, {Ye}, {Yu}, {Barlow}, {Bauer}, {Beck}, {Belicki}, {Biswas}, {Brinnel},
  {Brooke}, {Bue}, {Bulla}, {Burruss}, {Connolly}, {Cromer}, {Cunningham},
  {Dekany}, {Delacroix}, {Desai}, {Duev}, {Feeney}, {Flynn}, {Frederick},
  {Gal-Yam}, {Giomi}, {Groom}, {Hacopians}, {Hale}, {Helou}, {Henning},
  {Hover}, {Hillenbrand}, {Howell}, {Hung}, {Imel}, {Ip}, {Jackson}, {Kaspi},
  {Kaye}, {Kowalski}, {Kramer}, {Kuhn}, {Landry}, {Laher}, {Mao}, {Masci},
  {Monkewitz}, {Murphy}, {Nordin}, {Patterson}, {Penprase}, {Porter},
  {Rebbapragada}, {Reiley}, {Riddle}, {Rigault}, {Rodriguez}, {Rusholme}, {van
  Santen}, {Shupe}, {Smith}, {Soumagnac}, {Stein}, {Surace}, {Szkody}, {Terek},
  {Van Sistine}, {van Velzen}, {Vestrand}, {Walters}, {Ward}, {Zhang}, \&
  {Zolkower}}]{graham2019}
{Graham}, M.~J., {Kulkarni}, S.~R., {Bellm}, E.~C., {et~al.} 2019, \pasp, 131,
  078001, \dodoi{10.1088/1538-3873/ab006c}

\bibitem[{{Haiman} {et~al.}(2009){Haiman}, {Kocsis}, \&
  {Menou}}]{2009ApJ...700.1952H}
{Haiman}, Z., {Kocsis}, B., \& {Menou}, K. 2009, \apj, 700, 1952,
  \dodoi{10.1088/0004-637X/700/2/1952}

\bibitem[{{Holgado} {et~al.}(2018){Holgado}, {Sesana}, {Sandrinelli}, {Covino},
  {Treves}, {Liu}, \& {Ricker}}]{2018MNRAS.481L..74H}
{Holgado}, A.~M., {Sesana}, A., {Sandrinelli}, A., {et~al.} 2018, \mnras, 481,
  L74, \dodoi{10.1093/mnrasl/sly158}

\bibitem[{{Homan} {et~al.}(2021){Homan}, {Cohen}, {Hovatta}, {Kellermann},
  {Kovalev}, {Lister}, {Popkov}, {Pushkarev}, {Ros}, \&
  {Savolainen}}]{2021arXiv210904977H}
{Homan}, D.~C., {Cohen}, M.~H., {Hovatta}, T., {et~al.} 2021, arXiv e-prints,
  arXiv:2109.04977.
\newblock \doarXiv{2109.04977}

\bibitem[{{Hovatta} {et~al.}(2012){Hovatta}, {Lister}, {Aller}, {Aller},
  {Homan}, {Kovalev}, {Pushkarev}, \& {Savolainen}}]{2012AJ....144..105H}
{Hovatta}, T., {Lister}, M.~L., {Aller}, M.~F., {et~al.} 2012, \aj, 144, 105,
  \dodoi{10.1088/0004-6256/144/4/105}

\bibitem[{{Kellermann} {et~al.}(1998){Kellermann}, {Vermeulen}, {Zensus}, \&
  {Cohen}}]{1998AJ....115.1295K}
{Kellermann}, K.~I., {Vermeulen}, R.~C., {Zensus}, J.~A., \& {Cohen}, M.~H.
  1998, \aj, 115, 1295, \dodoi{10.1086/300308}

\bibitem[{{Komatsu} {et~al.}(2009){Komatsu}, {Dunkley}, {Nolta}, {Bennett},
  {Gold}, {Hinshaw}, {Jarosik}, {Larson}, {Limon}, {Page}, {Spergel},
  {Halpern}, {Hill}, {Kogut}, {Meyer}, {Tucker}, {Weiland}, {Wollack}, \&
  {Wright}}]{2009ApJS..180..330K}
{Komatsu}, E., {Dunkley}, J., {Nolta}, M.~R., {et~al.} 2009, \apjs, 180, 330,
  \dodoi{10.1088/0067-0049/180/2/330}

\bibitem[{{Liodakis} {et~al.}(2017){Liodakis}, {Pavlidou}, {Hovatta},
  {Max-Moerbeck}, {Pearson}, {Richards}, \& {Readhead}}]{2017MNRAS.467.4565L}
{Liodakis}, I., {Pavlidou}, V., {Hovatta}, T., {et~al.} 2017, \mnras, 467,
  4565, \dodoi{10.1093/mnras/stx432}

\bibitem[{{Lister} {et~al.}(2018){Lister}, {Aller}, {Aller}, {Hodge}, {Homan},
  {Kovalev}, {Pushkarev}, \& {Savolainen}}]{2018ApJS..234...12L}
{Lister}, M.~L., {Aller}, M.~F., {Aller}, H.~D., {et~al.} 2018, \apjs, 234, 12,
  \dodoi{10.3847/1538-4365/aa9c44}

\bibitem[{{Lister} \& {Homan}(2005)}]{2005AJ....130.1389L}
{Lister}, M.~L., \& {Homan}, D.~C. 2005, \aj, 130, 1389, \dodoi{10.1086/432969}

\bibitem[{{Lister} {et~al.}(2019){Lister}, {Homan}, {Hovatta}, {Kellermann},
  {Kiehlmann}, {Kovalev}, {Max-Moerbeck}, {Pushkarev}, {Readhead}, {Ros}, \&
  {Savolainen}}]{2019ApJ...874...43L}
{Lister}, M.~L., {Homan}, D.~C., {Hovatta}, T., {et~al.} 2019, \apj, 874, 43,
  \dodoi{10.3847/1538-4357/ab08ee}

\bibitem[{{Lomb}(1976)}]{1976Ap&SS..39..447L}
{Lomb}, N.~R. 1976, \apss, 39, 447, \dodoi{10.1007/BF00648343}

\bibitem[{Maggiore(2008)}]{maggiore2008}
Maggiore, M. 2008, Gravitational Waves: Volume 1: Theory and Experiments,
  Gravitational Waves (OUP Oxford)

\bibitem[{{Masci} {et~al.}(2019){Masci}, {Laher}, {Rusholme}, {Shupe}, {Groom},
  {Surace}, {Jackson}, {Monkewitz}, {Beck}, {Flynn}, {Terek}, {Landry},
  {Hacopians}, {Desai}, {Howell}, {Brooke}, {Imel}, {Wachter}, {Ye}, {Lin},
  {Cenko}, {Cunningham}, {Rebbapragada}, {Bue}, {Miller}, {Mahabal}, {Bellm},
  {Patterson}, {Juri{\'c}}, {Golkhou}, {Ofek}, {Walters}, {Graham}, {Kasliwal},
  {Dekany}, {Kupfer}, {Burdge}, {Cannella}, {Barlow}, {Van Sistine}, {Giomi},
  {Fremling}, {Blagorodnova}, {Levitan}, {Riddle}, {Smith}, {Helou}, {Prince},
  \& {Kulkarni}}]{masci2019}
{Masci}, F.~J., {Laher}, R.~R., {Rusholme}, B., {et~al.} 2019, \pasp, 131,
  018003, \dodoi{10.1088/1538-3873/aae8ac}

\bibitem[{{Max-Moerbeck} {et~al.}(2014){Max-Moerbeck}, {Richards}, {Hovatta},
  {Pavlidou}, {Pearson}, \& {Readhead}}]{2014MNRAS.445..437M}
{Max-Moerbeck}, W., {Richards}, J.~L., {Hovatta}, T., {et~al.} 2014, \mnras,
  445, 437, \dodoi{10.1093/mnras/stu1707}

\bibitem[{{O'Dea} {et~al.}(1986){O'Dea}, {Dent}, {Kinzel}, \&
  {Balonek}}]{1986AJ.....92.1262O}
{O'Dea}, C.~P., {Dent}, W.~A., {Kinzel}, W.~M., \& {Balonek}, T.~J. 1986, \aj,
  92, 1262, \dodoi{10.1086/114260}

\bibitem[{{Owen} {et~al.}(1985){Owen}, {O'Dea}, {Inoue}, \&
  {Eilek}}]{1985ApJ...294L..85O}
{Owen}, F.~N., {O'Dea}, C.~P., {Inoue}, M., \& {Eilek}, J.~A. 1985, \apjl, 294,
  L85, \dodoi{10.1086/184514}

\bibitem[{{Perley}(2019)}]{perley2019}
{Perley}, D.~A. 2019, \pasp, 131, 084503, \dodoi{10.1088/1538-3873/ab215d}

\bibitem[{{Planck Collaboration} {et~al.}(2020){Planck Collaboration},
  {Aghanim}, {Akrami}, {Ashdown}, {Aumont}, {Baccigalupi}, {Ballardini},
  {Banday}, {Barreiro}, {Bartolo}, {Basak}, {Battye}, {Benabed}, {Bernard},
  {Bersanelli}, {Bielewicz}, {Bock}, {Bond}, {Borrill}, {Bouchet}, {Boulanger},
  {Bucher}, {Burigana}, {Butler}, {Calabrese}, {Cardoso}, {Carron},
  {Challinor}, {Chiang}, {Chluba}, {Colombo}, {Combet}, {Contreras}, {Crill},
  {Cuttaia}, {de Bernardis}, {de Zotti}, {Delabrouille}, {Delouis}, {Di
  Valentino}, {Diego}, {Dor{\'e}}, {Douspis}, {Ducout}, {Dupac}, {Dusini},
  {Efstathiou}, {Elsner}, {En{\ss}lin}, {Eriksen}, {Fantaye}, {Farhang},
  {Fergusson}, {Fernandez-Cobos}, {Finelli}, {Forastieri}, {Frailis},
  {Fraisse}, {Franceschi}, {Frolov}, {Galeotta}, {Galli}, {Ganga},
  {G{\'e}nova-Santos}, {Gerbino}, {Ghosh}, {Gonz{\'a}lez-Nuevo}, {G{\'o}rski},
  {Gratton}, {Gruppuso}, {Gudmundsson}, {Hamann}, {Handley}, {Hansen},
  {Herranz}, {Hildebrandt}, {Hivon}, {Huang}, {Jaffe}, {Jones}, {Karakci},
  {Keih{\"a}nen}, {Keskitalo}, {Kiiveri}, {Kim}, {Kisner}, {Knox},
  {Krachmalnicoff}, {Kunz}, {Kurki-Suonio}, {Lagache}, {Lamarre}, {Lasenby},
  {Lattanzi}, {Lawrence}, {Le Jeune}, {Lemos}, {Lesgourgues}, {Levrier},
  {Lewis}, {Liguori}, {Lilje}, {Lilley}, {Lindholm}, {L{\'o}pez-Caniego},
  {Lubin}, {Ma}, {Mac{\'\i}as-P{\'e}rez}, {Maggio}, {Maino}, {Mandolesi},
  {Mangilli}, {Marcos-Caballero}, {Maris}, {Martin}, {Martinelli},
  {Mart{\'\i}nez-Gonz{\'a}lez}, {Matarrese}, {Mauri}, {McEwen}, {Meinhold},
  {Melchiorri}, {Mennella}, {Migliaccio}, {Millea}, {Mitra},
  {Miville-Desch{\^e}nes}, {Molinari}, {Montier}, {Morgante}, {Moss}, {Natoli},
  {N{\o}rgaard-Nielsen}, {Pagano}, {Paoletti}, {Partridge}, {Patanchon},
  {Peiris}, {Perrotta}, {Pettorino}, {Piacentini}, {Polastri}, {Polenta},
  {Puget}, {Rachen}, {Reinecke}, {Remazeilles}, {Renzi}, {Rocha}, {Rosset},
  {Roudier}, {Rubi{\~n}o-Mart{\'\i}n}, {Ruiz-Granados}, {Salvati}, {Sandri},
  {Savelainen}, {Scott}, {Shellard}, {Sirignano}, {Sirri}, {Spencer},
  {Sunyaev}, {Suur-Uski}, {Tauber}, {Tavagnacco}, {Tenti}, {Toffolatti},
  {Tomasi}, {Trombetti}, {Valenziano}, {Valiviita}, {Van Tent}, {Vibert},
  {Vielva}, {Villa}, {Vittorio}, {Wandelt}, {Wehus}, {White}, {White},
  {Zacchei}, \& {Zonca}}]{2020A&A...641A...6P}
{Planck Collaboration}, {Aghanim}, N., {Akrami}, Y., {et~al.} 2020, \aap, 641,
  A6, \dodoi{10.1051/0004-6361/201833910}

\bibitem[{Ragazzini \& Zadeh(1952)}]{ragazzini1952analysis}
Ragazzini, J.~R., \& Zadeh, L.~A. 1952, Transactions of the American Institute
  of Electrical Engineers, Part II: Applications and Industry, 71, 225

\bibitem[{{Readhead} {et~al.}(1989){Readhead}, {Lawrence}, {Myers}, {Sargent},
  {Hardebeck}, \& {Moffet}}]{1989ApJ...346..566R}
{Readhead}, A.~C.~S., {Lawrence}, C.~R., {Myers}, S.~T., {et~al.} 1989, \apj,
  346, 566, \dodoi{10.1086/168039}

\bibitem[{{Rector} \& {Stocke}(2001)}]{2001AJ....122..565R}
{Rector}, T.~A., \& {Stocke}, J.~T. 2001, \aj, 122, 565, \dodoi{10.1086/321179}

\bibitem[{{Rector} \& {Stocke}(2003)}]{2003AJ....125.2447R}
---. 2003, \aj, 125, 2447, \dodoi{10.1086/374768}

\bibitem[{{Ren} {et~al.}(2021){Ren}, {Ding}, {Zhang}, {Xue}, {Zhang}, {Xiong},
  {Li}, \& {Li}}]{2021MNRAS.506.3791R}
{Ren}, G.-W., {Ding}, N., {Zhang}, X., {et~al.} 2021, \mnras, 506, 3791,
  \dodoi{10.1093/mnras/stab1739}

\bibitem[{{Richards} {et~al.}(2011){Richards}, {Max-Moerbeck}, {Pavlidou},
  {King}, {Pearson}, {Readhead}, {Reeves}, {Shepherd}, {Stevenson},
  {Weintraub}, {Fuhrmann}, {Angelakis}, {Zensus}, {Healey}, {Romani}, {Shaw},
  {Grainge}, {Birkinshaw}, {Lancaster}, {Worrall}, {Taylor}, {Cotter}, \&
  {Bustos}}]{2011ApJS..194...29R}
{Richards}, J.~L., {Max-Moerbeck}, W., {Pavlidou}, V., {et~al.} 2011, \apjs,
  194, 29, \dodoi{10.1088/0067-0049/194/2/29}

\bibitem[{{Rodriguez} {et~al.}(2006){Rodriguez}, {Taylor}, {Zavala}, {Peck},
  {Pollack}, \& {Romani}}]{2006ApJ...646...49R}
{Rodriguez}, C., {Taylor}, G.~B., {Zavala}, R.~T., {et~al.} 2006, \apj, 646,
  49, \dodoi{10.1086/504825}

\bibitem[{{Sandrinelli} {et~al.}(2018){Sandrinelli}, {Covino}, {Treves},
  {Holgado}, {Sesana}, {Lindfors}, \& {Ramazani}}]{2018A&A...615A.118S}
{Sandrinelli}, A., {Covino}, S., {Treves}, A., {et~al.} 2018, \aap, 615, A118,
  \dodoi{10.1051/0004-6361/201732550}

\bibitem[{{Sandrinelli} {et~al.}(2017){Sandrinelli}, {Covino}, {Treves},
  {Lindfors}, {Raiteri}, {Nilsson}, {Takalo}, {Reinthal}, {Berdyugin}, {Fallah
  Ramazani}, {Kadenius}, {Tuominen}, {Kehusmaa}, {Bachev}, \&
  {Strigachev}}]{2017A&A...600A.132S}
---. 2017, \aap, 600, A132, \dodoi{10.1051/0004-6361/201630288}

\bibitem[{{Sbarufatti} {et~al.}(2006){Sbarufatti}, {Treves}, {Falomo}, {Heidt},
  {Kotilainen}, \& {Scarpa}}]{2006AJ....132....1S}
{Sbarufatti}, B., {Treves}, A., {Falomo}, R., {et~al.} 2006, \aj, 132, 1,
  \dodoi{10.1086/503031}

\bibitem[{{Scargle}(1982)}]{1982ApJ...263..835S}
{Scargle}, J.~D. 1982, \apj, 263, 835, \dodoi{10.1086/160554}

\bibitem[{{Scheuer} \& {Readhead}(1979)}]{1979Natur.277..182S}
{Scheuer}, P.~A.~G., \& {Readhead}, A.~C.~S. 1979, \nat, 277, 182,
  \dodoi{10.1038/277182a0}

\bibitem[{{Schwarzenberg-Czerny}(1989)}]{czerny1989}
{Schwarzenberg-Czerny}, A. 1989, \mnras, 241, 153,
  \dodoi{10.1093/mnras/241.2.153}

\bibitem[{{Shen} {et~al.}(2021){Shen}, {Chen}, {Hwang}, {Liu}, {Zakamska},
  {Oguri}, {Li}, {Lazio}, \& {Breiding}}]{2021NatAs...5..569S}
{Shen}, Y., {Chen}, Y.-C., {Hwang}, H.-C., {et~al.} 2021, Nature Astronomy, 5,
  569, \dodoi{10.1038/s41550-021-01323-1}

\bibitem[{{Sillanpaa} {et~al.}(1988){Sillanpaa}, {Haarala}, {Valtonen},
  {Sundelius}, \& {Byrd}}]{1988ApJ...325..628S}
{Sillanpaa}, A., {Haarala}, S., {Valtonen}, M.~J., {Sundelius}, B., \& {Byrd},
  G.~G. 1988, \apj, 325, 628, \dodoi{10.1086/166033}

\bibitem[{{Sobacchi} {et~al.}(2017){Sobacchi}, {Sormani}, \&
  {Stamerra}}]{2017MNRAS.465..161S}
{Sobacchi}, E., {Sormani}, M.~C., \& {Stamerra}, A. 2017, \mnras, 465, 161,
  \dodoi{10.1093/mnras/stw2684}

\bibitem[{{Teraesranta} {et~al.}(1998){Teraesranta}, {Tornikoski}, {Mujunen},
  {Karlamaa}, {Valtonen}, {Henelius}, {Urpo}, {Lainela}, {Pursimo}, {Nilsson},
  {Wiren}, {Laehteenmaeki}, {Korpi}, {Rekola}, {Heinaemaeki}, {Hanski},
  {Nurmi}, {Kokkonen}, {Keinaenen}, {Joutsamo}, {Oksanen}, {Pietilae},
  {Valtaoja}, {Valtonen}, \& {Koenoenen}}]{1998A&AS..132..305T}
{Teraesranta}, H., {Tornikoski}, M., {Mujunen}, A., {et~al.} 1998, \aaps, 132,
  305, \dodoi{10.1051/aas:1998297}

\bibitem[{{Uttley} {et~al.}(2002){Uttley}, {McHardy}, \&
  {Papadakis}}]{2002MNRAS.332..231U}
{Uttley}, P., {McHardy}, I.~M., \& {Papadakis}, I.~E. 2002, \mnras, 332, 231,
  \dodoi{10.1046/j.1365-8711.2002.05298.x}

\bibitem[{{Valtonen} {et~al.}(2016){Valtonen}, {Zola}, {Ciprini}, {Gopakumar},
  {Matsumoto}, {Sadakane}, {Kidger}, {Gazeas}, {Nilsson}, {Berdyugin},
  {Piirola}, {Jermak}, {Baliyan}, {Alicavus}, {Boyd}, {Campas Torrent},
  {Campos}, {Carrillo G{\'o}mez}, {Caton}, {Chavushyan}, {Dalessio}, {Debski},
  {Dimitrov}, {Drozdz}, {Er}, {Erdem}, {Escartin P{\'e}rez}, {Fallah Ramazani},
  {Filippenko}, {Ganesh}, {Garcia}, {G{\'o}mez Pinilla}, {Gopinathan},
  {Haislip}, {Hudec}, {Hurst}, {Ivarsen}, {Jelinek}, {Joshi}, {Kagitani},
  {Kaur}, {Keel}, {LaCluyze}, {Lee}, {Lindfors}, {Lozano de Haro}, {Moore},
  {Mugrauer}, {Naves Nogues}, {Neely}, {Nelson}, {Ogloza}, {Okano}, {Pandey},
  {Perri}, {Pihajoki}, {Poyner}, {Provencal}, {Pursimo}, {Raj}, {Reichart},
  {Reinthal}, {Sadegi}, {Sakanoi}, {Salto Gonz{\'a}lez}, {Sameer}, {Schweyer},
  {Siwak}, {Sold{\'a}n Alfaro}, {Sonbas}, {Steele}, {Stocke}, {Strobl},
  {Takalo}, {Tomov}, {Tremosa Espasa}, {Valdes}, {Valero P{\'e}rez},
  {Verrecchia}, {Webb}, {Yoneda}, {Zejmo}, {Zheng}, {Telting}, {Saario},
  {Reynolds}, {Kvammen}, {Gafton}, {Karjalainen}, {Harmanen}, \&
  {Blay}}]{2016ApJ...819L..37V}
{Valtonen}, M.~J., {Zola}, S., {Ciprini}, S., {et~al.} 2016, \apjl, 819, L37,
  \dodoi{10.3847/2041-8205/819/2/L37}

\bibitem[{{Vaughan} {et~al.}(2016){Vaughan}, {Uttley}, {Markowitz},
  {Huppenkothen}, {Middleton}, {Alston}, {Scargle}, \&
  {Farr}}]{2016MNRAS.461.3145V}
{Vaughan}, S., {Uttley}, P., {Markowitz}, A.~G., {et~al.} 2016, \mnras, 461,
  3145, \dodoi{10.1093/mnras/stw1412}

\bibitem[{{Welsh}(1999)}]{1999PASP..111.1347W}
{Welsh}, W.~F. 1999, \pasp, 111, 1347, \dodoi{10.1086/316457}

\bibitem[{{Xin} {et~al.}(2021){Xin}, {Mingarelli}, \&
  {Hazboun}}]{2021ApJ...915...97X}
{Xin}, C., {Mingarelli}, C. M.~F., \& {Hazboun}, J.~S. 2021, \apj, 915, 97,
  \dodoi{10.3847/1538-4357/ac01c5}

\bibitem[{{Zechmeister} \& {K{\"u}rster}(2009)}]{2009A&A...496..577Z}
{Zechmeister}, M., \& {K{\"u}rster}, M. 2009, \aap, 496, 577,
  \dodoi{10.1051/0004-6361:200811296}

\end{thebibliography}
\bibliographystyle{aasjournal}

\end{document}